\documentclass[aps,prd,reprint,twocolumn,preprintnumbers,floatfix,nofootinbib]{revtex4-1}

\usepackage[utf8]{inputenc}
\usepackage[dvips]{graphicx}
\usepackage[dvipsnames]{xcolor}
\usepackage{color}
\usepackage{relsize}
\usepackage{graphics}
\usepackage{epstopdf}
\usepackage{hyperref}
\usepackage{mathrsfs}
\usepackage{amssymb}
\usepackage{physics}
\usepackage{booktabs}
\renewcommand{\figurename}{Fig.}
\usepackage[normalem]{ ulem }
\usepackage{amsthm}
\usepackage{amsmath}
\usepackage{cancel}
\usepackage{float}
\usepackage{fontawesome} 
\usepackage[caption=false]{subfig}
\usepackage{tabularx,ragged2e} 
 \newcolumntype{L}{>{\RaggedRight\arraybackslash}X}

\usepackage{hyperref}

\newcommand{\be}{\begin{equation}}
\newcommand{\ee}{\end{equation}}
\newcommand{\bea}{\begin{eqnarray}}
\newcommand{\eea}{\end{eqnarray}}
\newcommand{\ba}{\begin{aligned}}
\newcommand{\ea}{\end{aligned}}

\newcommand{\mDM}{m_{\rm DM}}

\newcommand{\TBH}{T_{\rm BH}}
\newcommand{\MBH}{M_{\rm BH}}

\newcommand{\MBHi}{M_{\rm BH}^{\rm in}}
\newcommand{\MPL}{M_p}

\newcommand{\Neff}{N_{\rm eff}}
\newcommand{\DNeff}{\Delta N_{\rm eff}}

\newcommand{\figref}[1]{Fig.~\ref{#1}}

\newcommand{\secref}[1]{Section~\ref{#1}}

\newcommand\myshade{80}
\colorlet{mylinkcolor}{ForestGreen}
\colorlet{mycitecolor}{Red}
\colorlet{myurlcolor}{violet}

\hypersetup{
 linkcolor = mylinkcolor!\myshade!black,
 citecolor = mycitecolor!\myshade!black,
 urlcolor = myurlcolor!\myshade!black,
 colorlinks = true
}

\usepackage{pgfplots}
\newcommand{\equaref}[1]{Eq.~(\ref{#1})}

\pgfplotsset{compat=1.17}

\begin{document}
\sloppy  

\preprint{IPPP/22/49}

\vspace*{1mm}

\title{Redshift Effects in Particle Production from Kerr Primordial Black Holes}

\author{Andrew Cheek$^{a}$}
\email{acheek@camk.edu.pl}
\author{Lucien Heurtier$^{b}$}
\email{lucien.heurtier@durham.ac.uk}
\author{Yuber F. Perez-Gonzalez$^{b}$}
\email{yuber.f.perez-gonzalez@durham.ac.uk}
\author{Jessica Turner$^{b}$}
\email{jessica.turner@durham.ac.uk}

\affiliation{$^a$ Astrocent, Nicolaus Copernicus Astronomical Center of the Polish Academy of Sciences, ul.Rektorska 4, 00-614 Warsaw, Poland}
\affiliation{$^b$ Institute for Particle Physics Phenomenology, Durham University, South Road, Durham, U.K.}

\begin{abstract} 
When rotating primordial black holes evaporate via Hawking radiation, their rotational energy and mass are dissipated with different dynamics. 
We investigate the effect of these dynamics on the production of dark radiation -- in the form of hot gravitons or vector bosons -- and non-cold dark matter. 
Although the production of higher-spin particles is enhanced while primordial black holes are rotating, we show that the energy density of dark radiation experiences an extra redshift because their emission effectively halts before PBH evaporation completes.
We find that taking this effect into account leads to suppression by a factor of $\mathcal{O}(10)$ of $\Delta N_{\rm eff}$ for maximally rotating black holes as compared to previous results. 
Using the solution of the Friedmann and Boltzmann equations to accurately calculate the evolution of linear perturbations, we revisit the warm dark matter constraints for light candidates produced by evaporation and how these limits vary over black hole spins.
Due to the interplay of enhanced production and late dilution, we obtain that higher spin particles are most affected by these bounds.
Our code {\tt FRISBHEE}, FRIedmann Solver for Black Hole Evaporation in the Early universe, developed for this work can be found at
 \href{https://github.com/yfperezg/frisbhee}{\faGithub}.
\end{abstract}

\maketitle

\section{Introduction}

Determining the evolution of the early Universe before Big-Bang Nucleosynthesis (BBN) era is not an easy task. Although observational data regarding the Cosmic Microwave Background (CMB) may constrain very early stages of the Universe evolution such as cosmic inflation, very little is known about the post-inflationary phase of our Universe's history. Indeed, the thermalization of Standard Model (SM) particles at early time is expected to erase all traces of the initial conditions which led to the hot big bang era we are familiar with. The discovery of Gravitational Waves (GWs)~\cite{LIGOScientific:2016aoc,LIGOScientific:2017vwq} opened the door to new ways of probing the evolution of the early Universe, either through the detection of recent astrophysical events or through the measurement of a stochastic background of GWs produced in the primordial Universe (see, e.g.~Ref.~\cite{LISACosmologyWorkingGroup:2022jok} and references therein for a review).  
Historically, however, measurements of the number of effective relativistic degrees of freedom $\Neff$ has provided important information regarding the Universe before BBN, for example, by confirming the description of neutrino decoupling. In the realm of physics beyond the SM, measurements of $\Neff$ provide vital constraints on models with light particles.
Many different models predict a deviation of $\Neff$ from the minimal situation where only neutrinos and photons dominate the Standard Model (SM) radiation energy density, see e.g.~Refs.~\cite{Jeong:2013eza,Buen-Abad:2015ova,Lesgourgues:2015wza,Buen-Abad:2017gxg,Blinov:2020hmc,Chacko:2003dt,Chacko:2004cz,Berlin:2017ftj}.

A particularly appealing scenario that has attracted renewed attention is the formation and evaporation of Primordial Black Holes (PBHs)~\cite{Zeldovich:1967lct,Hawking:1971ei,Carr:1974nx}.
PBHs could have dominated the evolution of the Universe before the BBN epoch, leading to observational tests such as GWs~\cite{Inomata:2020lmk,Domenech:2020ssp} or contributions to $\Neff$ as PBHs emit all existing degrees of freedom in nature~\cite{Hawking:1974rv, Hawking:1974sw} including light states that could modify the number of relativistic degrees of freedom, see Ref.~\cite{Auffinger:2022khh} for a recent and complete review. Among the many Beyond the Standard Model (BSM) light particles predicted in diverse scenarios, one candidate stands out, the massless spin-2 two mediator of Gravity, the \emph{graviton}. As many consider the graviton's existence to be guaranteed, one could, in principle, constrain the presence of PBHs in the pre-BBN Early Universe by measuring either modification to $\Neff$, parameterized by $\DNeff$, or GWs in future experiments~\cite{Dong:2015yjs}. However, let us note that if the PBHs were of the Schwarzschild type, graviton emission is suppressed compared to other degrees of freedom.
The rationale behind such diminished production lies in the dependence of the Hawking emission rate on absorption probabilities, or Graybody Factors, which are reduced at low energies for larger values of the total angular momentum of the incoming --- or emitted --- particle.

If, on the other hand, PBHs had an initial non-zero angular momentum, the emission of higher spin particles becomes enhanced~\cite{Page:1976df,Page:1976ki,Page:1977um}, leading to an augmented number density of emitted gravitons, which in turn could lead to observable modifications to $\Neff$, as first studied in Refs.~\cite{Hooper:2019gtx,Hooper:2020evu,Masina:2020xhk,Masina:2021zpu,Arbey:2021ysg}.
These works computed the contribution to $\DNeff$ considering the time evolution of both mass and spin and demonstrated that future CMB-HD~\cite{CMB-HD:2022bsz} experiments could, in principle, probe the existence of PBHs. 
In this paper, we revisit the derivation of $\DNeff$ for hot gravitons produced by the evaporation of spinning --- Kerr --- PBHs.
Since the BH angular momentum is depleted much faster than its mass, the enhanced emission of spin-2 particles only occurs in the first stages of evaporation, such that graviton production effectively stops before the PBHs have entirely evaporated.
Thus, there is an additional and crucial redshift that gravitons experience, which, in the past has been neglected or partially accounted for, when calculating $\DNeff$
We thus improve on previous treatments by carefully tracking the time evolution of PBH mass and angular momentum, together with the Universe's evolution.
We assume a monochromatic mass and spin distribution and consider that the PBH population acquired a significant angular momentum via some unspecified mechanism.
We find a difference of a factor of $\mathcal{O}(10)$ to previous estimates for close-to-maximally rotating PBHs in the determination of $\DNeff$.
Moreover, we determine the sensitivity to the initial PBH density fraction as a function of the initial PBH mass in the range $10^{-1}\lesssim \MBH^{\rm in} (~{\rm g}) \lesssim 10^{9}$ for gravitons and new light vectors from future measurements of $\DNeff$.

The unique dynamics that Kerr PBHs possess will also produce important modifications in the generation of additional and stable BSM states, which could constitute the dark matter (DM)~\cite{Fujita:2014hha,Lennon:2017tqq, Morrison:2018xla,Hooper:2019gtx, Masina:2020xhk, Baldes:2020nuv, Gondolo:2020uqv, Bernal:2020kse, Bernal:2020ili,Auffinger:2020afu, Bernal:2020bjf,Arbey:2021ysg, Masina:2021zpu, JyotiDas:2021shi, Cheek:2021odj, Cheek:2021cfe, Sandick:2021gew,Bernal:2022oha}.
Constraints from structure formation, especially for light DM, where $m_{\rm DM }\lesssim 10^3\,\rm{GeV}$, significantly impact the allowed parameter space, as demonstrated in~\cite{Baldes:2020nuv,Auffinger:2020afu, Masina:2021zpu}.
In the second part of this work, we return to the derivation of such limits on light DM by solving the complete set of Friedmann-Boltzmann equations together with angular momentum and mass depletion rates to provide accurate inputs for calculating the evolution of linear perturbations in the Universe. For the latter task, we use {\tt CLASS} \cite{CLASSI, CLASSII, CLASSIV} to obtain updated warm dark matter constraints. Moreover, we perform a systematic study of how these constraints vary over black hole spins, finding that higher spin particles are most affected, where the same story of enhanced production and late time dilution plays out, but with the additional effect of redshift acting to slow the matter particles.

This manuscript is organized as follows.
In \secref{sec:KerrPBH}, we describe the time evolution of Kerr PBHs, emphasizing the angular moment evaporation.
Next, we detail the determination of $\DNeff$ for gravitons and additional light vectors in \secref{sec:deltaN}, comparing with existing results in the literature. Finally, \secref{sec:darkmatter} is devoted to the specific scenario of light DM emission from Kerr PBHs, where we determine the redshifted phase-space distribution of DM and the subsequent rederivation of hot DM constraints.
We draw our conclusions in \secref{sec:concls}.
Throughout, we assume natural units where $\hbar = c = k_{\rm B} = 1$.

\section{Evaporation of Kerr PBHs}\label{sec:KerrPBH}

When PBHs evaporate, they emit particles of all kinds at rates that differ depending on the particle's quantum numbers and the mass and angular momentum of the evaporating PBH. 
As consequence, their mass and angular momentum are depleted at rates that depend on the initial PBH properties.
To derive such evaporation equations, let us consider the Hawking emission rate of any particle species $i$, which within a time ($\dd t$) and momentum ($\dd p$) interval, given by~\cite{Hawking:1974rv,Hawking:1974sw}
\begin{align}\label{eq:KBHrate}
\frac{\dd^2 \mathcal{N}_{i}}{\dd p\dd t}&=\frac{g_i}{2\pi^2} \sum_{l=s_i}\sum_{m=-l}^l\frac{\dd^2 \mathcal{N}_{ilm}}{\dd p\,\dd t}\,,
\end{align}
with
\begin{align}\label{eq:KBHrate_lm}
\frac{\dd^2 \mathcal{N}_{ilm}}{\dd p\dd t}&=\frac{\sigma_{s_i}^{lm}(\MBH,p,a_\star)}{\exp\left[(E_i - m\Omega)/\TBH\right]-( -1)^{2s_i}}\frac{p^3}{E_i}\,,
\end{align}
where $\MBH$ is the instantaneous PBH mass, $E_i$ ($g_i$) is the energy (internal degrees of freedom) of particle $i$, $\Omega = (a_\star/2G\MBH)(1/(1+\sqrt{1-a_\star^2}))$ is the angular velocity of the horizon,  $G$ is Newton's constant, and $l,m$ are the total and axial angular momentum quantum numbers, respectively.
An essential quantity in Eq.~\eqref{eq:KBHrate_lm} is the absorption cross-section, $\sigma_{s_i}^{lm}$, which characterizes the possible back-scattering of particles in the presence of centrifugal and gravitational potentials~\cite{Hawking:1974rv,Hawking:1974sw,Page:1976df,Page:1977um}.
Such cross-sections are determined from reflection and transmission coefficients of scatterings by the gravitational potential after solving equations of motions in a curved spacetime around the BH.
For the specific case of spinning BHs, we have followed the method established in Refs.~\cite{Chandrasekhar:1975zz,Chandrasekhar:1976zz,Chandrasekhar:1977kf}, where the problem is reduced to solve Schr\"odinger-like one-dimensional wave equations whose potentials depend on the particle's spin.

The BH mass and spin loss rates are calculated by summing Eq.~\eqref{eq:KBHrate} over the different species and integrating over the phase space, to obtain \cite{PhysRevD.41.3052,PhysRevD.44.376}
\begin{align}\label{eq:dynamicsevaporation}
 \frac{d\MBH}{dt} &= - \epsilon(\MBH, a_\star)\frac{M_p^4}{\MBH^2}\,,\\
 \label{eq:dynamicsspin}\frac{da_\star}{dt} &= - a_\star[\gamma(\MBH, a_\star) - 2\epsilon(\MBH, a_\star)]\frac{M_p^4}{\MBH^3}\,,
\end{align}
where $M_p=1.22\times10^{19}\,{\rm GeV}$ denotes the Planck mass, $a_\star = J M_{\rm Pl}^2/\MBH^2
$ is the dimensionless spin parameter of the BH, and $\epsilon(\MBH, a_\star)$ ($\gamma(\MBH, a_\star)$) is the evaporation (angular momentum evaporation) function which depends on the BH instantaneous mass and spin and the total degrees of freedom existing in nature.
For a given particle type $i$, the functions $\gamma_i(\MBH,a_\star)$ and $\varepsilon_i(\MBH,a_\star)$ are 
\begin{align}
 \varepsilon_i(\MBH, a_\star) &= \frac{g_i}{2\pi^2}\int_{0}^\infty \sum_{l=s_i}\sum_{m=-l}^l\frac{\dd^2 \mathcal{N}_{ilm}}{\dd p\dd t}\,E dE\,,\\
 \gamma_i(\MBH, a_\star) &= \frac{g_i}{2\pi^2}\int_{0}^\infty \sum_{l=s_i}\sum_{m=-l}^l m \frac{\dd^2 \mathcal{N}_{ilm}}{\dd p\dd t}\, dE\,,
\end{align}
where the details on how to calculate these parameters can be found in Ref.~\cite{Cheek:2021odj}.
\begin{figure}[t!]
 \def\sepf{0.52}
	\centering
 \includegraphics[width=0.45\textwidth]{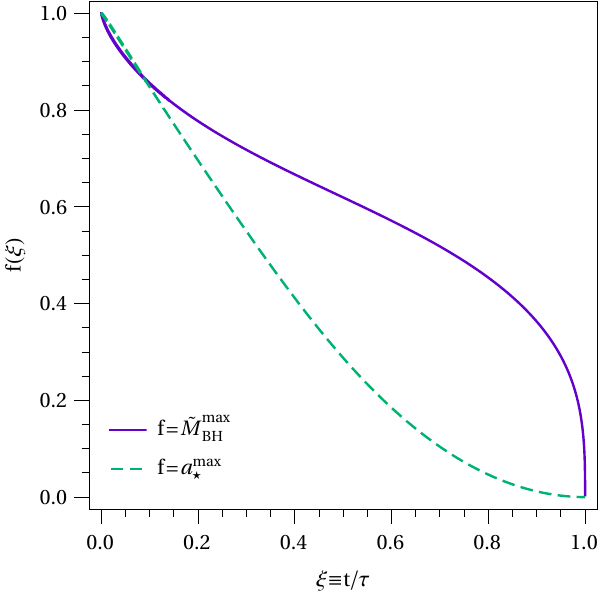}
 \caption{The purple curve shows the evolution of $\widetilde{M}_{\rm BH}^{\rm max}(\xi)\equiv\MBH(\xi)/\MBH^{\rm in}$, the BH mass normalised to initial mass while the green dashed curve presents the angular momentum depletion $a_\star^{\rm max}(\xi)$ as a function $\xi$, the time normalised to the BH total lifetime. The presented solutions are obtained for an initially maximally rotating BH ($a_{\star,\,\rm in}=1$).
 The time evolution for any other $a_{\star,\,\rm in}<1$ can be derived from these curves after performing a suitable transformation, see text.}
 \label{fig:massandspinloss}
\end{figure} 

Let us now consider the time evolution of a close-to-maximally rotating BH, having
an initial spin parameter $a_{\rm\star, in}\to 1$ and an initial mass $\MBHi$.
In \figref{fig:massandspinloss}, we present the numerical solution to the system of equations where $f(\xi)$ denotes alternatively $\widetilde{M}_{\rm BH}^{\rm max}
(t)\equiv\MBH(\xi)/\MBH^{\rm in}$ (purple) or $a_\star^{\rm max}(\xi)$ (green dashed), being $\xi\equiv t/\tau\in[0,1]$, with $\tau$ denoting the PBH lifetime.
As seen from this figure, the spin of a Kerr BH is generally depleted via Hawking evaporation much faster than the mass. 
Indeed, at around $t\sim 0.65\tau$, one can observe that the spin has dropped over one order of magnitude ($a_\star \sim 0.1$), whereas the mass has only decreased by a half ($M \sim 0.5 M_{\rm in}$). 
As we will see in the proceeding sections, this rapid decrease of the spin, compared to the mass, implies that a careful treatment of Hawking evaporation, which simultaneously takes into account the evolution of the Universe, is required when dealing with a particle production that is strongly spin-dependent. 

Interestingly, it was shown in Ref.~\cite{Page:1976ki} that, in the limit of light evaporation products, the solution of Eqs.~\eqref{eq:dynamicsevaporation}-\eqref{eq:dynamicsspin} can be generically obtained for any initial mass and spin from the evolution of a close-to-maximally rotating BH, $a_{\rm\star, in}=1$. 
To obtain the generic solution, the following transformations must be performed:
\begin{subequations}
 \begin{align}
  a_\star(\xi)&=a_\star^{\rm max}([1-x_i]\xi+x_i)\,,\\
  \widetilde{M}_{\rm BH}(\xi)&=\frac{\widetilde{M}_{\rm BH}^{\rm max}([1-x_i]\xi+x_i)}{\widetilde{M}_{\rm BH}^{\rm max}(x_i)}\,,
 \end{align}
\end{subequations}
where $x_i$ is such that $a_\star^{\rm max}(x_i)=a_{\star,\,\rm in}$. 
We then can extend the previous discussion about the time dependence of higher spin particle emission.
For the case where $a_{\star,\,\rm in}=0.7$, average spin value from hierarchical merger rates~\cite{Fishbach:2017dwv}, we find that only $\sim 40\%$ of the initial mass has been depleted when $a_\star = 0.1$, demonstrating that the importance of correctly tracking the evolution of the PBH mass and spin with respect of the overall evolution of the Universe.

\section{Determination of $\Delta N_{\rm eff}$} \label{sec:deltaN}
The number of relativistic degrees of freedom in the early Universe is extremely sensitive to the thermal history of the Universe's particle physics content. All SM particles, including neutrinos, are in thermal equilibrium at high temperatures. As the temperature of the thermal plasma decreases, SM particles become non-relativistic and decoupled from the thermal bath, releasing their energy to lighter species through annihilation or decay processes. Neutrinos, however, decouple while relativistic, effectively following a Fermi-Dirac distribution, despite their interaction rate becoming inefficient compared to the Universe's expansion rate. Assuming that the neutrinos decoupled instantaneously, the subsequent decoupling of the electrons from the thermal bath is expected to inject entropy exclusively in the photon bath, which effectively shifts the neutrino temperature as compared with the photon temperature:
\be
T_\nu = (4/11)^{1/3}T_\gamma\,.
\ee
In the absence of any further considerations, one would expect the energy density of radiation today to follow the relation
\be\label{eq:Neff_SM}
\rho_\mathrm{R}^\mathrm{SM}=\rho_\gamma\left[1+\frac{7}{8}\left(\frac{T_\nu}{T_\gamma}\right)N_\mathrm{eff}^{\rm SM}\right]\,,
\ee
where $\rho_\gamma$ is the energy density of the photon bath and $N_\mathrm{eff}^{\rm SM}$ is the number of relativistic neutrinos, which would naively be $3$. However, this is in fact $N_\mathrm{eff}\approx 3.045$~\cite{Mangano:2001iu, deSalas:2016ztq}. The reason for this extra contribution is twofold: $(i)$ electrons transfer a fraction of entropy into the neutrino sector through QED corrections, and, most importantly, $(ii)$ the decoupling of neutrinos is {\em not} instantaneous. The amount of entropy transferred from the electron sector to neutrinos out-of-equilibrium is temperature-dependent. Therefore, the non-instantaneous nature of neutrino decoupling is critical as it affects how much entropy the electrons can transfer to the neutrino sector after decoupling. 

In the event that a dark sector may contribute to the energy density of radiation today, one usually defines this extra amount of relativistic degrees of freedom, $\Delta N_\mathrm{eff}$, as
\be
\rho_\mathrm{R}\equiv \rho_\gamma\left[1+\frac{7}{8}\left(\frac{T_\nu}{T_\gamma}\right)(N_\mathrm{eff}^{\rm SM}+\Delta N_\mathrm{eff})\right]\,.
\ee
Denoting the energy density of dark radiation as $\rho_\mathrm{DR}$ and using Eq.~\eqref{eq:Neff_SM} together with the fact that $\rho_\mathrm{R}= \rho_\mathrm{R}^\mathrm{SM}+\rho_\mathrm{DR}$ allows to invert this relation to obtain
\be
\Delta N_{\rm eff} \equiv \left\{\frac{8}{7}\left(\frac{4}{11}\right)^{-\frac{4}{3}}+N_{\rm eff}^{\rm SM}\right\} 
 \frac{\rho_{\rm DR}}{\rho_{\rm R}^{\rm SM}}\,,
\ee
where both $\rho_\mathrm{R}^\mathrm{SM}$ and $\rho_\mathrm{DR}$ are evaluated at present time. Computing the contribution of PBH evaporation to $\Delta N_\mathrm{eff}$ therefore requires the precise evaluation of the energy density of dark radiation, $\rho_\mathrm{R}$, produced during the evaporation process.

The possibility that PBHs contribute significantly to $\Delta N_\mathrm{eff}$ was thoroughly studied by different groups \cite{Masina:2020xhk,Masina:2021zpu,Hooper:2020evu,Arbey:2021ysg, Hooper:2019gtx}. Those studies showed that the Hawking production of particles with spin $\geqslant 1$ becomes significantly enhanced in the presence of a large PBH spin. In the context of dark matter production from PBH evaporation, this same conclusion was reached in Refs.~\cite{Cheek:2021odj, Cheek:2021cfe} in a complete numerical analysis, including the full grey body factors and the dynamical solving of Boltzmann and Friedman equations.

By nature, the evaporation of PBHs is {\em not} an instantaneous process and, similarly to the decoupling of neutrinos in the SM, it is natural to question whether the time evolution of the evaporation process for Kerr PBHs may affect predictions for the value of $\Delta N_\mathrm{eff}$. As we have seen in \secref{sec:KerrPBH}, the dynamics of the evaporation are encoded in a two-dimensional differential system, describing the non-trivial evolution of both the mass and spin of the BH as a function of time (see Eq.~\eqref{eq:dynamicsevaporation} and \eqref{eq:dynamicsspin}). Spinning PBHs tend to lose most of their spin in the first phase of the evaporation before the remaining mass evaporates in the second phase. These complicated dynamics have two major effects: $(i)$ the total lifetime of a spinning PBH is affected by the initial value of its spin, and $(ii)$ the production of vector and tensor particles is significantly enhanced during the first phase of the evaporation (when the spin is non-zero) but becomes similar to the one of a Schwarzschild PBH towards the end of the evaporation process. In Refs.~\cite{Arbey:2021ysg, Hooper:2020evu, Hooper:2019gtx} the first of those two effects was considered via calculation of the PBH lifetime in the presence of the BH spin. However, instantaneous evaporation was assumed when computing different particle abundances, affecting the contribution to $\Delta N_\mathrm{eff}$. To go beyond this approximation, we solve the Friedmann equations simultaneously with the evaporation dynamical system of Eq.~\eqref{eq:dynamicsevaporation} and \eqref{eq:dynamicsspin}. We will assume PBHs were formed after overdense perturbations entered the Hubble horizon during a radiation-dominated era. This common assumption leads to the generic one-to-one relation between the initial mass of the PBH and the initial energy density of the Universe at the time of PBH formation
\begin{align}\label{eq:Min}
 \MBH^{\rm in} = \frac{4\pi}{3} \gamma\frac{\rho_i}{H_i^3}\,,
\end{align}
where $\gamma=0.2$ which is a dimensionless gravitational collapse parameter and $H_i$ is the Hubble parameter~\cite{Carr:2020gox}.
 For further convenience, we will quantify the energy fraction of PBHs in the Universe, at the time of their formation, using the parameter
\begin{align}\label{eq:betap}
 \beta^\prime \equiv \gamma^{1/2}\left(\frac{g_\star (T_{\rm in})}{106.75}\right)^{-1/4}\frac{\rho^{\rm in}_{\rm PBH}}{\rho^{\rm in}}\,.
\end{align}
The set of equations that are to be solved numerically consists of the Friedmann equation
\be\label{eq:Hubble}
\frac{3H^2M_p^2}{8\pi}=\rho_\mathrm{R}^{\rm SM} + \rho_{\rm DR} + \rho_{\rm PBH}\,,
\ee
together with the set of Boltzmann equations describing the evolution of the SM radiation (R), dark radiation (DR) and PBH energy densities throughout the history of the Universe are respectively given by
\cite{Gutierrez:2017ibk,Cheek:2021odj}
\begin{subequations}\label{eq:FBEqs}
\begin{align}
 \dot\rho_\mathrm{R}^{\rm SM} + 4H \rho_\mathrm{R}^{\rm SM} &= -\left.\frac{\dd \log\MBH}{\dd t}\right|_{\rm SM}\rho_{\rm PBH}\,,\\
 \dot{\rho}_{\rm DR} + 4H \rho_{\rm DR} &= -\left.\frac{\dd \log\MBH}{\dd t}\right|_{\rm DR}\rho_{\rm PBH}\,,\\
 \dot{\rho}_{\rm PBH} + 3H \rho_{\rm PBH} &= \frac{\dd \log\MBH}{\dd t}\rho_{\rm PBH}\,,
\end{align}
\end{subequations}
respectively.
{We determine the solutions of the Friedmann, Boltzmann and BH evolution equations using our code called FRIedmann Solver for Black Hole Evaporation in the Early universe, {\tt FRISBHEE}, which will be publicly available in \href{https://github.com/yfperezg/frisbhee}{\faGithub}\footnote{A detailed instruction manual for the usage of {\tt FRISBHEE} will be released in the near future.}.}
\begin{figure*}[t]
 \centering
 \includegraphics[width=0.9\linewidth]{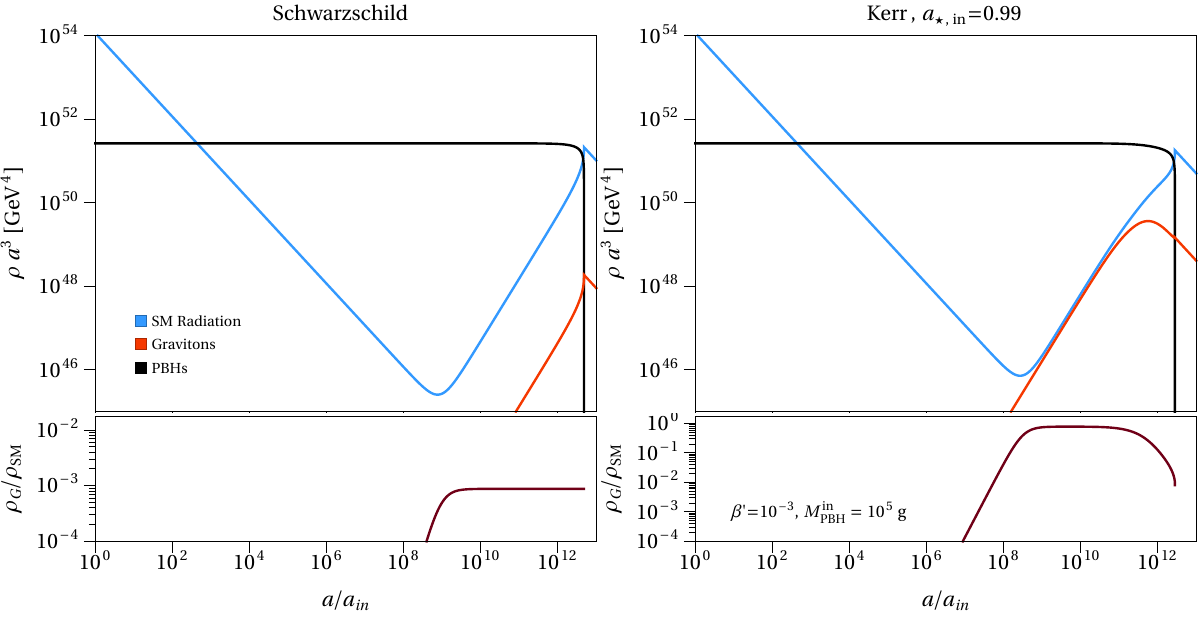}
 \caption{\label{fig:examsol} Evolution of the energy density of PBH (blue) and SM radiation (black) in the case of Schwarzschild black holes (left) and Kerr black holes with spin $a_\star = 0.99$ (right). The bottom panels depict the evolution of the graviton contribution to the total energy density of radiation during the evaporation process.}
\end{figure*}

Once the evaporation has completed, the SM evolves independently from DR, and the energy densities of the two sectors at present time can easily be extrapolated from their values after evaporation using entropy conservation. The value of $\Delta N_\mathrm{eff}$ can therefore be calculated using 
\cite{Hooper:2019gtx}
\begin{widetext}
\begin{align}\label{eq:Neffg}
 \Delta N_{\rm eff} = \left\{\frac{8}{7}\left(\frac{4}{11}\right)^{-\frac{4}{3}}+N_{\rm eff}^{\rm SM}\right\} 
 \frac{\rho_{\rm DR}(T_{\rm ev})}{\rho_{\rm R}^{\rm SM}(T_{\rm ev})}
 \left(\frac{g_*(T_{\rm ev})}{g_*(T_{\rm eq})}\right)
 \left(\frac{g_{*S}(T_{\rm eq})}{g_{*S}(T_{\rm ev})}\right)^{\frac{4}{3}}\,,
\end{align}
\end{widetext}
where $N_{\rm eff}^{\rm SM} = 3.045$ denotes the effective number of relativistic neutrinos in the SM \cite{deSalas:2016ztq}, $T_{\rm ev}$ is the plasma temperature when the PBHs have evaporated, $T_{\rm eq}=0.75$ eV is the temperature at which matter-radiation equality occurs.  

\subsection{Graviton contributions to $\Delta N_{{\rm eff}}$}

Despite having no hopes of detecting the graviton anytime soon, its existence is often assumed when quantum field theorists consider gravitation. As any massless spin-2 field would give rise to gravity, this further cements the belief that such a particle is likely to exist. Given the graviton's status as an `honorary' member of the Standard Model, one can view the constraints on its energy density produced by evaporating BHs as a pure test of the possible PBH abundance in the early Universe. \\

To understand the effect of redshifting during the non-instantaneous evaporation, we present in Fig.~\ref{fig:examsol} an example of the evolution of the energy densities for a PBH with $\MBHi=10^5$ g and assuming $\beta^\prime = 10^{-3}$.
In the left (right) panel, we show the evolution of the energy density
of the SM radiation and PBHs in blue and black, respectively, for a Schwarzschild (Kerr with $a_\star=0.99$) PBHs. In the bottom panels, the energy density of the gravitons
relative to the SM radiation is shown. In the Schwarzschild case, the PBHs come to dominate the energy density of the Universe at $a\sim 10^{3}$ and their evaporation becomes relevant at $a\sim 10^{12}$; however, the SM radiation energy density begins to increase around $a\sim 10^{9}$. We see that the fraction of the radiation energy density consisting of gravitons increases simultaneously as the visible radiation energy density increases and then plateaus as the gravitons contribute a constant fraction towards the total radiation energy density. For Kerr PBHs, we observe that gravitons are amply produced while the PBHs spins are non-zero as the fractional energy density of the gravitons to the total radiation increases sharply (is overall larger than in the Schwarzschild case) and plateaus. However, in the final stage of evaporation, when the rotational energy has been depleted, the PBH remains massive and continues to evaporate.
Meanwhile, the gravitons are free-streaming thus only experiencing redshift. Therefore, the graviton contribution to the total radiation bath is reduced as photon production dominates (shown by the steep decrease after $\log_{10}(a)\sim11$ in the lower panel).

\begin{figure*}[t!]
 \includegraphics[width=0.7\linewidth]{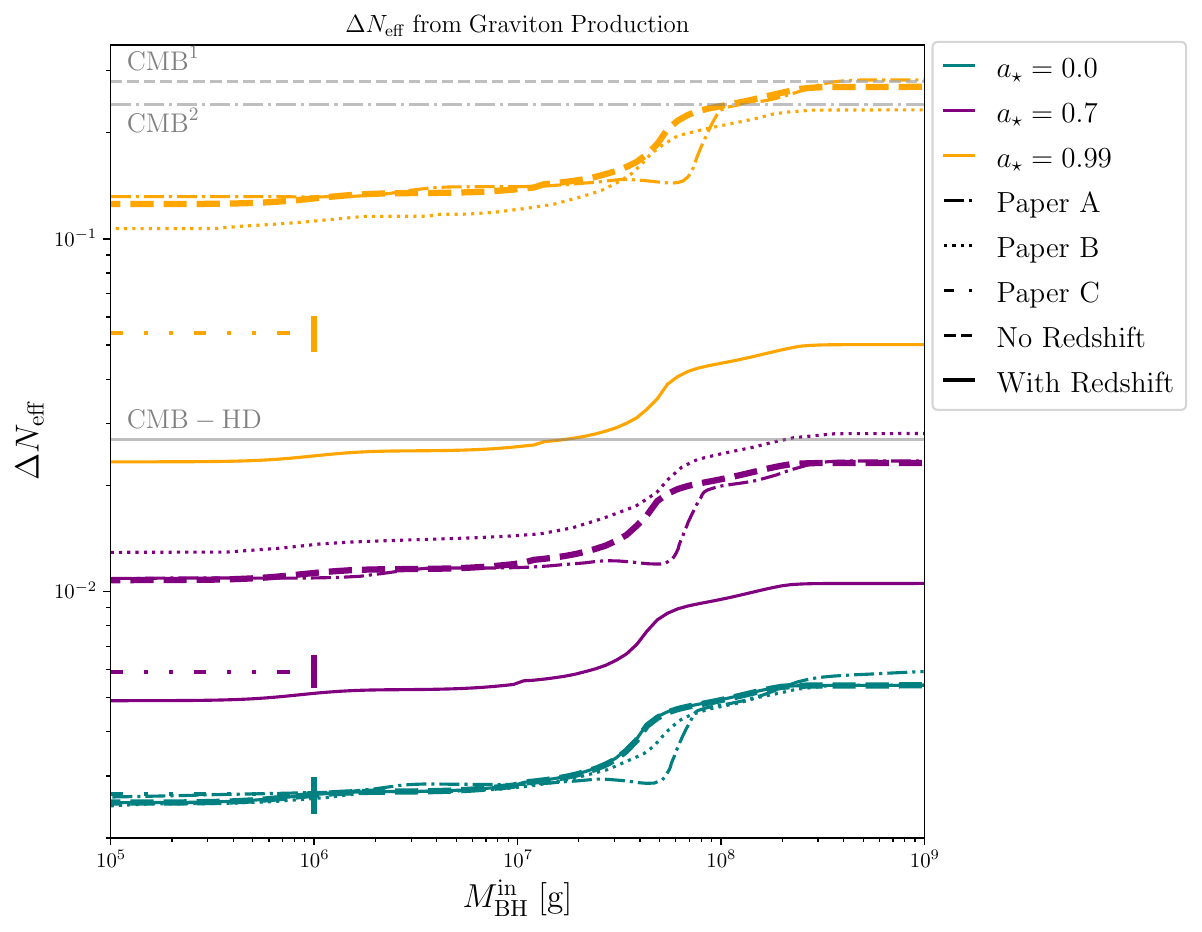}
 \caption{\label{fig:DNeff} PBH contribution to $\Delta N_{\rm eff}$ assuming only the Standard Model and a massless graviton. We assume PBH domination and monochromatic BH distributions in mass $M_{\rm in}$ and spin $a_{\star}$. The solid lines correspond to our calculation, solving for the number densities in a time evolving manner. The teal, purple and orange lines correspond to $a_{\star}=0.0,0.7$ and $0.99$ respectively. We overlay sensitivities of current and future experimental sensitivities to $\Delta N_{\rm eff}$ at $2\sigma$ C.L. Current limits of \cite{Planck:2018vyg} and are denoted  as CMB$^1$ (TT + low E, conservative) and CMB$^2$ (TT,TE,EE + low E, more stringent) and future projections from CMB-HD~\cite{CMB-HD:2022bsz}. We also compare our results to Ref.~\cite{Hooper:2020evu},  Ref.~\cite{Arbey:2021ysg}, and Ref.~\cite{Masina:2021zpu} referred to as Paper A, Paper B, and Paper C, respectively.
Also included are the results to our calculation neglecting the effects of redshift from noninstantaneous BH evaporation, see text for details.}
\end{figure*}
In Fig.~\ref{fig:DNeff}, we show the predictions for $\Delta N_{\rm eff}$ that we obtain when primordial black holes of various spins dominate the energy density of the Universe evaporate and produce massless gravitons as they evaporate (solid lines). The heavier the initial PBH mass, the more significant the contribution to $\Delta N_{\rm eff}$ as heavier PBHs have longer lifetimes and produce more gravitons. Furthermore, we observe that the higher the initial spin of the PBH populations, the more significant the contribution to the $\Delta N_{\rm eff}$ as the graviton production is enhanced. To visualize how the non-instantaneous nature of the evaporation affects the results, we also derived similar results where we account for the redshift of the evaporation products only after the PBHs entirely evaporated (dashed lines). As one can see, the results match in the case of Schwarzschild PBHs ($a_\star = 0$), whereas a discrepancy develops for increasing spins. For the case $a_\star=0.99$, we observe that the effect of including redshift {during the PBH evaporation process} reduces the contribution to $\Delta N_{\rm eff}$ by almost an order of magnitude over the entire considered mass range of PBHs when we compare our results (solid line) with the findings of Ref.~\cite{Hooper:2020evu} (dot-dashed line), Ref.~\cite{Arbey:2021ysg} (dotted line), and Ref.~\cite{Masina:2021zpu} (dot-dot-dashed horizontal bar) which we refer to as {\em Paper A}, {\em Paper B} and {\em Paper C}, respectively, in Fig.~\ref{fig:DNeff}. In {\em Paper B} it was noted that the dynamics of the evaporation are modified in the case of a Kerr black hole, which affects the value of $\rho_{\rm R}(T_{\rm EV})$ at the end of evaporation. However, it is assumed that the energy density of dark radiation at evaporation $\rho_{\rm DR}(T_{\rm EV})$ is affected in the same proportion, which led the authors of {\em Paper B} to claim that this does not affects the calculation of $\Delta N_{\rm eff}$ since only the ratio of the two quantities appears in Eq.~\eqref{eq:Neffg}. The authors of {\em Paper A} do not explicitly explain whether their calculation of $\Delta N_{\rm eff}$ includes the effect of the dark-radiation redshift, they obtain quantitatively similar results to {\em Paper B}, which suggests that they used a similar approximation. Interestingly, {\em Paper C} in the figure (horizontal dot-dot-dashed bar), does show a discrepancy with the former papers in the case of Kerr PBHs. This reference considers the redshift when integrating over the evaporation dynamics. However, {\em Paper C} also assumes that the Universe is exactly matter-dominated until PBHs fully evaporate. This does not fully account for the cosmological dynamics of the evaporation and explains why the results of {\em Paper C} are somewhat intermediate between our results and the results presented in {\em Paper A} and {\em Paper B}.

Unsurprisingly, our results which do not account for the redshift of the evaporation products during the evaporation process, match the results of {\em Paper A} and {\em Paper B} relatively well. This is easy to understand: the production of gravitons via PBH evaporation is enhanced for larger PBH spin. However, as we have seen in \secref{sec:KerrPBH}, the spin of a Kerr PBH is depleted by evaporation much faster than its mass. 
Gravitons, which are thus produced copiously at the beginning of the evaporation, behave like radiation, while the mass energy of the remaining PBHs behaves like matter until PBHs fully evaporate mostly into SM particles. In the case of PBHs with large spin, the time difference between the production of gravitons and SM particles is large. In that case the energy density of gravitons redshifts significantly until the final production of SM particles during which the production of gravitons is negligible. This has the effect of depleting the value of $\rho_{\rm DR}(T_{\rm EV})$ used in Eq.~\eqref{eq:Neffg}, which, in turn, explains the large suppression of $\Delta N_{\rm eff}$ that we obtain for non-zero PBH spins. This redshift effect can only be captured by solving both the mass and spin loss rate of the PBHs while simultaneously tracking in time the evolution of the Universe.
We note that our results and those of {\em Papers A} and {\em Papers B} all exhibit the characteristic bump at 
$M_{\rm{in}}=8\times 10^{7}\,$g (corresponding to a reheating temperature of $\sim\,100$ MeV), which manifests from the fact that the number of degrees of freedom is sensitive to the QCD equation of state. 

In Fig.~\ref{fig:DNeff},  we  show some current and future predicted sensitivities to $\Delta N_{\rm eff}$ by three relevant CMB surveys: the two current constraints are taken from the Planck Collaboration \cite{Planck:2018vyg} and are denoted  as CMB$^1$ (TT + low E, conservative) and CMB$^2$ (TT,TE,EE + low E, more stringent). The future CMB survey, CMB-HD.~\cite{CMB-HD:2022bsz}, shows an order of magnitude improvement in sensitivity to  $\Delta N_{\rm eff}$ at the $2\sigma$ C.L.

For a review of the sensitivities of other cosmic surveys including CORE \cite{CORE:2016npo}, DESI \cite{DESI:2016fyo} and EUCLID \cite{EUCLID:2011zbd} see Ref.~\cite{Lattanzi:2017ubx}.
We observe that including the effect of redshift bears a pessimistic result for current or future experiments. Nonetheless, it is essential to realize that current CMB results are not close to constraining early maximally spinning PBH populations through the emission of gravitons.
The range of $\MBHi$ in \figref{fig:DNeff} is $10^5-10^9\, {\rm g}$ because the driving factor behind changes in $\DNeff$ is the variation of $g_*$ and $g_{*S}$, which for the SM, is constant outside the plot. One can simply extrapolate the endpoints of the limits, however when one goes to lower $\MBHi$ values, the possibility to have PBH domination becomes limited.
We note that we have considered the value of $a_\star=0.99$ in order to compare with the literature.
Larger values of $a_\star$ would increase the value of $\DNeff$ since the Hawking temperature and particle emission is significantly enhanced for larger angular momenta.
We have verified that, for a close-to-maximally rotating PBH with $a_\star=0.9999$, the contribution to $\DNeff$ is increased by a factor of $\sim 15\%$ in comparison to the case of $a_\star=0.99$.

\begin{figure}
\centering
\includegraphics[width=0.99\linewidth]{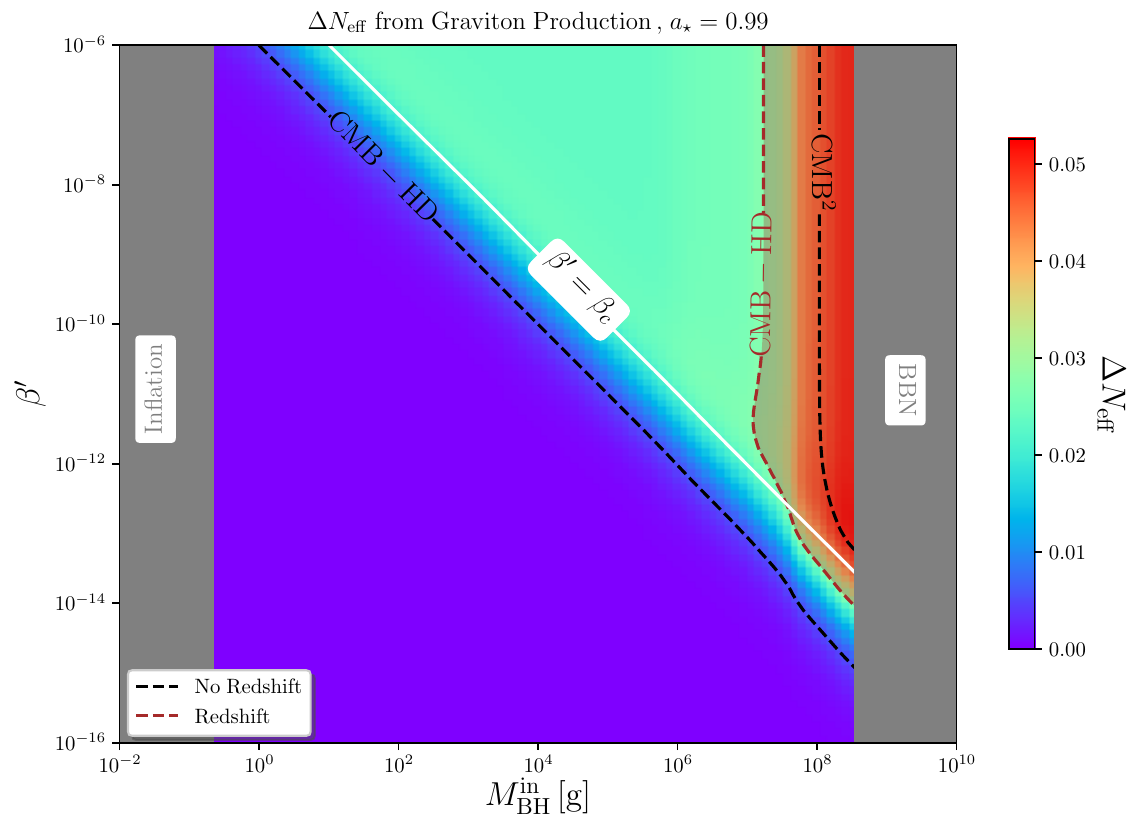}
\caption{\label{fig:graviton_a_099} Colour map showing PBH contribution to $\Delta N_{\rm eff}$ in the $\beta^\prime$ and $\MBHi$ plane. As in \figref{fig:DNeff} we assume only the Standard Model and massless graviton as well as monochromatic $\MBHi$ and $a_{\star}$ distributions. Here we show the results for $a_{\star}=0.99$. }
\end{figure}

In Fig.~\ref{fig:graviton_a_099}, we show the contribution of hot gravitons to $\Delta N_{\rm eff}$ from the evaporation of Kerr PBHs according to our calculation described above. 
The PBH distributions considered are monochromatic in $\MBHi$, which is varied along the x-axis, and we keep $a_{\star}=0.99$ fixed.
More realistic cases of mass and spin distributions will be investigated in forthcoming works.
We no longer assume that $\beta^\prime > \beta_c$, where $\beta_c$ is the critical value of the initial PBH density from which PBHs would eventually dominate the evolution of the Universe. 

One can see, that assuming instantaneous evaporation (black dashed line), CMB-HD would achieve sensitivity below $\beta_c$, which would constitute an even stronger limit than reported by Refs.~\cite{Hooper:2020evu, Arbey:2021ysg} because they always assume $\beta^\prime>\beta_c$. However, when the effect of redshift during the non-instantaneous evaporation is considered (brown dashed line), the projected sensitivity for CMB-HD is much less optimistic and the projected sensitivity for CMB-HD, if present, is above the $\beta_c$ except for when $\MBHi\gtrsim 10^7\,\,{\rm g}$.

A few comments on the shape of the limits are in order, we focus on the brown dashed line which more accurately captures the physics. In the $\beta^\prime> \beta_c$ region it is usually assumed that $\DNeff$ becomes insensitive to the precise value of $\beta$, because PBH domination has been achieved and one can effectively take this point as an initial condition. In this case then only $\MBHi$ is relevant as this determines when the injection of dark radiation happens and therefore, $\rho^{\rm SM}_{\rm R}$. This situation is certainly visible for $\beta^\prime \gtrsim 10^{-10}$ and we can say that PBHs totally dominate. Below this, the line starts to vary due to two effects, both of which enhance the contribution to $\DNeff$ as $\MBHi$ is increased. One effect comes with PBHs being less dominant, they therefore dilute $\rho_{\rm DR}$ less. The other comes from variations of $g_*$ and $g_{*S}$, which as seen in \equaref{eq:Neffg}, effects the value of $\DNeff$. For lower values still, where $\beta^\prime<\beta_c$, we do see that a small region of parameter space will be probed by CMB-HD, this is aided by the fact that $g_*$ has dropped. Showing that one need not assume PBH domination to hope to detect the effects on $\DNeff$. The shape of the limit here mimics the $1/\MBHi$ behaviour of the $\beta_c$ line because with a later evaporation time, a smaller PBH population has more time to exploit the difference in cosmic scaling. Finally, it can be noted that the production of gravitons from PBHs with spin $a=0.7$ escapes the sensitivity of future experiments, as can also be seen from Fig.~\ref{fig:DNeff}.

\subsection{BSM contributions to $\Delta N_{{\rm eff}}$}

This section explores possible contributions to $\Delta N_{\rm eff}$ from light particles, which are marginally more hypothetical than the graviton. However, with the recent determination of the muon anomalous magnetic moment, $a_\mu = (g-2)_\mu/2$, by the E989 experiment at Fermilab~\cite{Muong-2:2021ojo} and planned CERN NA64$\mu$~\cite{Gninenko:2640930,Gninenko:2653581} and Fermilab $M^3$~\cite{Kahn:2018cqs} experiments, the number of known light vectors may be about to change.  

As is well established, the emission of spin-1 particles is enhanced for rotating BHs as they attempt to shed their angular momentum. However, this enhancement is less pronounced than in the case of the graviton as they have a lower spin. On the other hand, the absorption cross-section, $\sigma_{s_i}$ as seen in Eq.~\eqref{eq:KBHrate}, calculated in the Schwarzchild metric for spin-1 particles, is much less suppressed than for spin-2 particles. This means that a more significant proportion of the radiation will be produced at late times, meaning they will not experience as much dilution as their spin-2 counterparts. 

\begin{figure*}
 \includegraphics[width=0.485\linewidth]{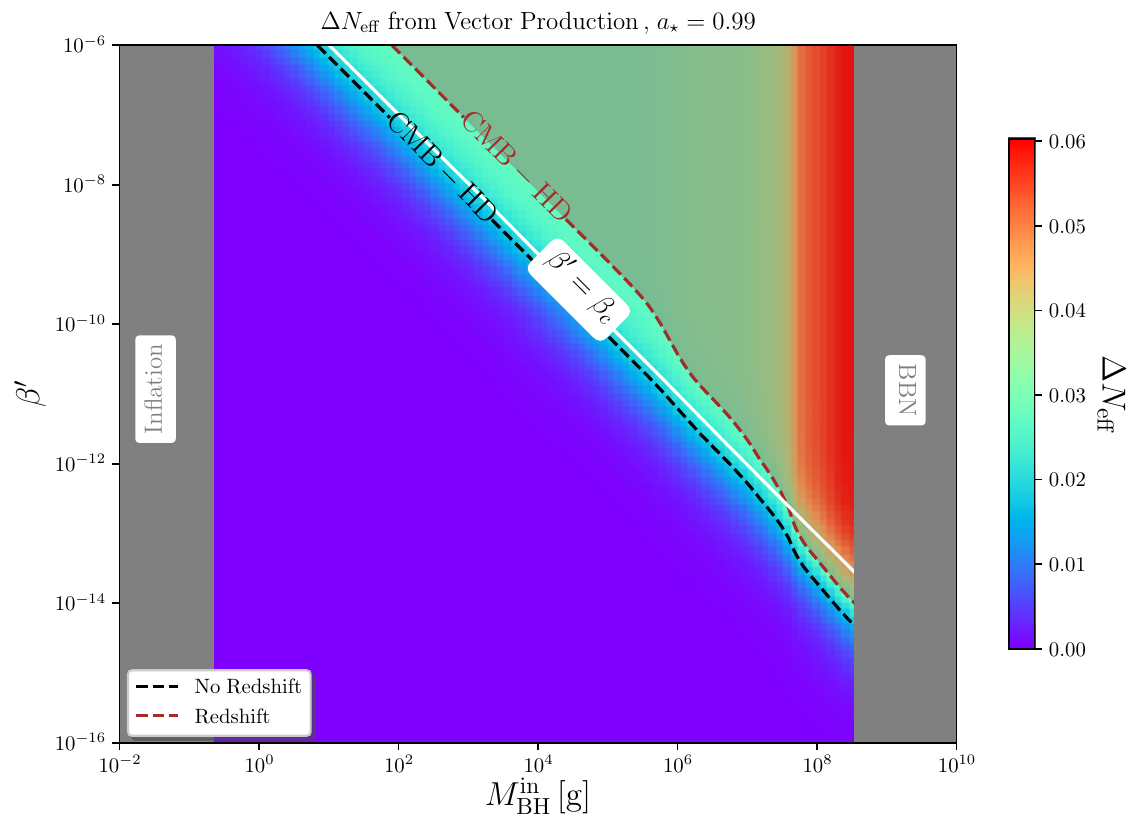}\hspace{0.03\linewidth}\includegraphics[width=0.485\linewidth]{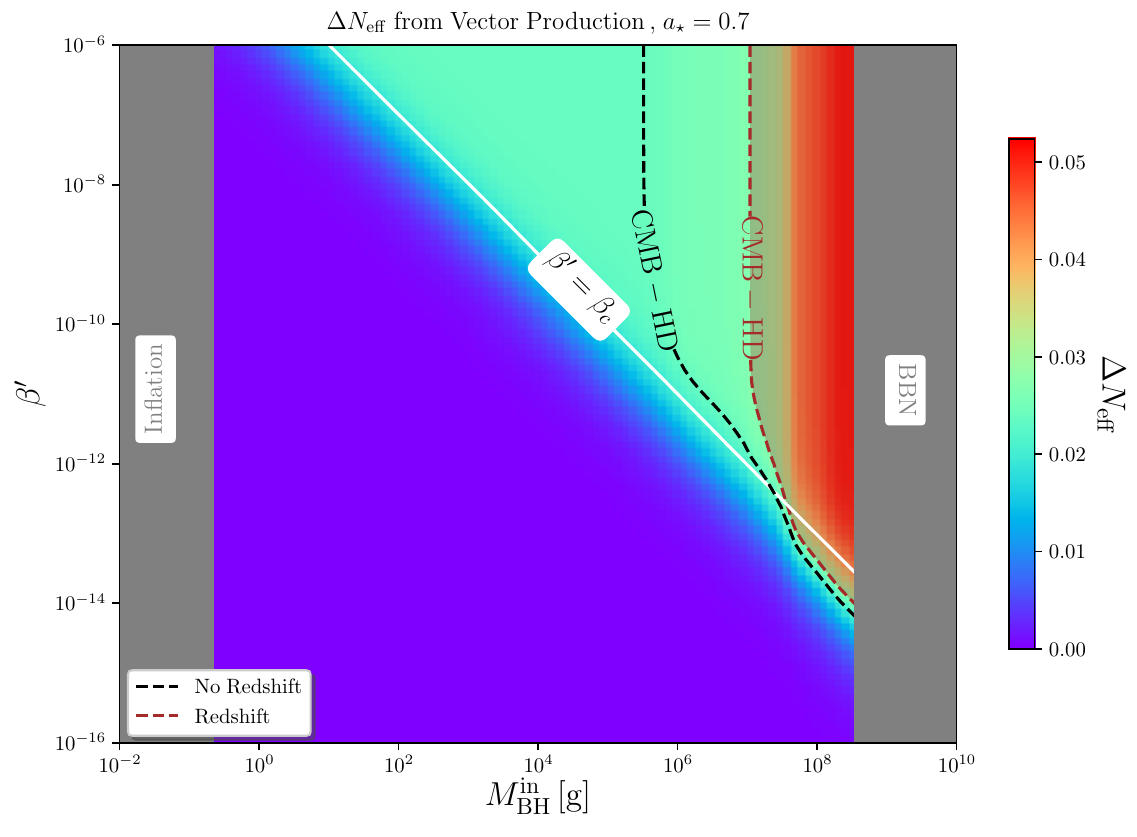}
 \caption{\label{fig:vector_a} Same as Fig.~\ref{fig:graviton_a_099}, but for a massless vector with two different values for BH spin $a_{\star}=0.99$ and $a_{\star}=0.7$ on the left and right panels respectively.}
\end{figure*}

In Fig.~\ref{fig:vector_a}, we show the effect this new vector would have on $\Delta N_{\rm eff}$ if there was a population of PBHs in the Early Universe. First, let us focus on the left panel and compare our results for the contribution of hot gravitons or dark vector bosons to $\Delta N_{\rm eff}$. Generally, the structure of these two plots resembles that of 
Fig.~\ref{fig:graviton_a_099}, where the contribution to $\Delta N_{\rm eff}$ become larger for greater initial energies densities of the PBH population; however, 
the picture looks less pessimistic. For both cases, taking the instantaneous evaporation approximation overestimates the sensitivity projections for CMB-HD, but to a much lesser extent. For $a_\star=0.99$, the projections when accounting for non-instantaneous evaporation is much more constraining than the tensor case in Fig.~\ref{fig:graviton_a_099}. Naturally the constraints from CMB-HD becomes weaker in the lower PBH spin case (right panel of Fig.~\ref{fig:vector_a}) than in the higher spin case (left panel of Fig.~\ref{fig:vector_a}) as less vector particles are produced as the initial PBH spin is lower. Furthermore, as we see in the right panel of Fig.~\ref{fig:vector_a}, CMB-HD will be sensitive to vector radiation from a population of black holes with a more modest spin, $a_\star=0.7$. In fact above $\MBHi\sim 10^8\,\,{\rm g}$ the sensitivity we estimate is very similar for both values of $a_\star$, where the lines go below $\beta_c$.

\section{Phase space distribution of matter}\label{sec:darkmatter}

If the emitted particles from PBHs are massive, stable and sufficiently inert, they may constitute part or all of the dark matter required by our current understanding of structure formation in the Universe. This mechanism of gravitational production of dark matter does not rely on any other interaction and is extremely efficient, particularly for heavy dark matter masses, up to $\mDM\sim\MPL$, see Refs.~\cite{Fujita:2014hha,Lennon:2017tqq, Hooper:2019gtx, Masina:2020xhk, Baldes:2020nuv, Gondolo:2020uqv, Bernal:2020kse, Bernal:2020ili,Auffinger:2020afu, Bernal:2020bjf,Arbey:2021ysg, Masina:2021zpu, JyotiDas:2021shi, Cheek:2021odj, Cheek:2021cfe, Sandick:2021gew,Bernal:2022oha} for further details. If PBH evaporation is solely responsible for producing massive non-interacting dark matter, then the dark matter would cool with the expansion of the Universe and be consistent with the cold dark matter picture. This may not necessarily be the case for masses below $~10^3$ GeV, and if the dark matter was produced too late, their free-streaming length might be incompatible with observations of small-scale structures ~\cite{Bode:2000gq}. 

In this section, we improve on our previous estimates for the warm dark matter constraints given in Ref.~\cite{Cheek:2021odj} by following the methodology of \cite{Baldes:2020nuv,Auffinger:2020afu, Masina:2021zpu}. The key difference from our previous work is that in this paper, we track the complete evolution of the dark matter phase-space distribution to obtain the input parameters for the computational tool for calculating Anisotropies, {\tt CLASS}~\cite{CLASSI, CLASSII, CLASSIV}. Furthermore, what separates this work from the literature Refs.~\cite{Baldes:2020nuv,Auffinger:2020afu, Masina:2021zpu}, is that we have numerically solved the Friedmann and Boltzmann equations and by using the precise evolution of the plasma temperature $T(t)$ and scale factor $a(t)$.\\

Following \cite{Baldes:2020nuv}, one needs to determine the phase space of the dark matter distribution, $f_{\rm DM}$ which is defined as, 
\begin{equation}
g_{\rm DM}f_{\rm DM}=\frac{{\rm d}n_{\rm DM}}{{\rm d}^3p},
\end{equation}
where $g_{\rm DM}$ is the number of degrees of freedom, $p$ is the three-momentum, and $n_{\rm DM}$ is the dark matter number density. One provides $f_{\rm DM}$ at the time of dark matter production to {\tt CLASS} . In our case this is $t_{\rm ev}$, which is easily defined for the monochromatic distributions. Therefore, one can write 
\begin{equation}
 f_{\rm DM}= \left.\frac{n_{\rm BH}\left(t_{\rm in}\right)}{g_{\rm DM}}\left(\frac{a(t_{\rm in})}{a(t)}\right)^3
\frac{1}{p^2}\frac{{\rm d}\mathcal{N}_{\rm DM}}{{\rm d}p}\right\vert_{t=t_{\rm ev}},
\label{eq:DMphasespace}
\end{equation}
where $t_{\rm in}$ is the scale factor at formation, and we have used the fact that the co-moving number density of BHs is constant up until they evaporate. The particle emission rate per momentum is calculated by taking the emission rate given in Eq.~\eqref{eq:KBHrate} and integrating over time,
\begin{equation}
 \frac{{\rm d}\mathcal{N}_{\rm DM}}{{\rm d}p}= \int_0^\tau {\rm d} t^\prime\frac{a(\tau)}{a(t')}\times\frac{{\rm d}^2\mathcal{N}_{\rm DM}}{{\rm d}p^\prime {\rm d } t^{\prime}}\left(p\frac{a(\tau)}{a(t')}, t'\right)\,,
 \label{eq:dNdP_redshift}
\end{equation}
one can see that the non-instantaneous nature of particle emission is taken into account by including the ratio of scale factors at a given $t$. Since the above determination of ${\rm d}\mathcal{N}_{\rm DM}/{\rm d}p$  is the norm in the literature, we do not expect differences on the scale found in \secref{sec:deltaN}, however we note that, as pointed out in Ref.~\cite{Cheek:2021odj}, the relic density calculation is highly sensitive to the value of $a(t)$ and $T_{\rm ev}$. In fact, $T_{\rm ev}$ also enters in the determination of the matter power spectrum as outlined in Ref.~\cite{Baldes:2020nuv}, when one evaluates the time-independent NCDM temperature,
\begin{equation}
 \mathcal{T}_{\rm ncdm}=T_{\rm in}\frac{a(t_{\rm ev})}{T(t_0)}=\frac{T_{\rm in}}{T_{\rm ev}}\left(\frac{g_{s\star}(T_0)}{g_{s\star}(T_{\rm ev})}\right)^{1/3}\,,
 \label{eq:tncdm}
\end{equation}
where $T_{\rm in}$ is the plasma temperature at the time of the PBH formation, which we assume to take place during radiation domination\footnote{
We can simply determine this initial temperature from the initial PBH mass in Eq.~\eqref{eq:Min}.
}, and $T(t_0)$ is the temperature today.\\

Equipped with the inputs from Eqs.~\eqref{eq:dNdP_redshift} and \eqref{eq:tncdm} one can determine the matter power spectrum, $P(K)$, from {\tt CLASS}, and compare to that of the cold dark matter (CDM), matter power spectrum by way of the transfer function, 
\begin{equation}
 P(k)=P_{\rm CDM}(k)T^2(k)\,,
 \label{eq:transfer_func}
\end{equation}
where $k$ is the wavenumber. Barring inconclusive or short lived anomalies, the CDM paradigm is largely supported by astrophysical observation~\cite{Fairbairn:2022gar} meaning that $T^2(k)\approx 1$ in the regions where experimental observations have been made. Naturally, observing structures at greater $k$ is increasingly difficult because such small galaxies do not support star formation. Measurements from Lyman-$\alpha$ are the probes of $P(k)$ at highest $k$~\cite{PhysRevD.88.043502,Palanque-Delabrouille:2019iyz}, see Refs.~\cite{Varma:2020kbq, He:2020rkj} for work on proposed studies at higher $k$. Following Refs.~\cite{Bode:2000gq,Viel:2005qj, Baldes:2020nuv} we perform parameter fits in the parameterisation 
\begin{equation}
 T(k)=(1+(\alpha k)^{2\mu})^{-5/\mu}\,,
 \label{eq:Transfer_param}
\end{equation}
where $\mu$ is dimensionless exponent which is fixed to $\mu=1.12$ as in Ref.~\cite{Viel:2005qj} and $\alpha$ is the breaking scale, which we take to be saturated at $\alpha=1.3\times 10^{-2}\,{\rm Mpc}\,h^{-1}$~\cite{Viel:2005qj, Palanque-Delabrouille:2019iyz, Garzilli:2019qki, Baldes:2020nuv}.\\

The above methodology has been performed previously in the literature. For example, in Refs.~\cite{Baldes:2020nuv, Auffinger:2020afu} these constraints were calculated for Schwarzschild black holes where Ref.~\cite{Auffinger:2020afu} uses emission rates shown in Eq.~\eqref{eq:KBHrate} from the {\tt BlackHawk} code. Ref.~\cite{Masina:2021zpu} estimated the effects Kerr distributions would have on the warm dark matter constraints by using a weighted integral of the emission rate. The primary difference in our work is that we use the solution of the coupled Boltzmann equations such that we have fully tracked the evolution of the scale factor and plasma temperature. For scalar, fermion and vector particles we use emission functions calculated following the procedure in Refs.~\cite{Chandrasekhar:1975zz,10.2307/79115,Chandrasekhar:1977kf} to compute the cross-sections $\sigma_{s_i}^{lm}$, where we find good agreement with {\tt BlackHawk}~\cite{Arbey:2019mbc}. For the spin-2 case \cite{Chandrasekhar:1976zz} we use the greybody factors from {\tt BlackHawk}~\cite{Arbey:2019mbc} which we checked to be consistent with Ref.~\cite{Dong:2015yjs}.\\

Using {\tt FRISBHEE}, for a given $m_{\rm DM}$, $a_\star$ and intrinsic particle spin, $s_{\rm DM}$, we determine the values of $\beta^\prime$ and $\MBHi$ such that the dark matter relic abundance matches that of the Planck collaboration's measured value, $\Omega_{\rm DM}\,h^2=0.12$~\cite{Planck:2018vyg}. We then take those parameter values to calculate $f_{\rm DM}$ and $\mathcal{T}_{\rm ncdm}$ using Eq.~\eqref{eq:DMphasespace} and Eq.~\eqref{eq:tncdm}.
\begin{figure}
 \centering
 \includegraphics[width=0.45\textwidth]{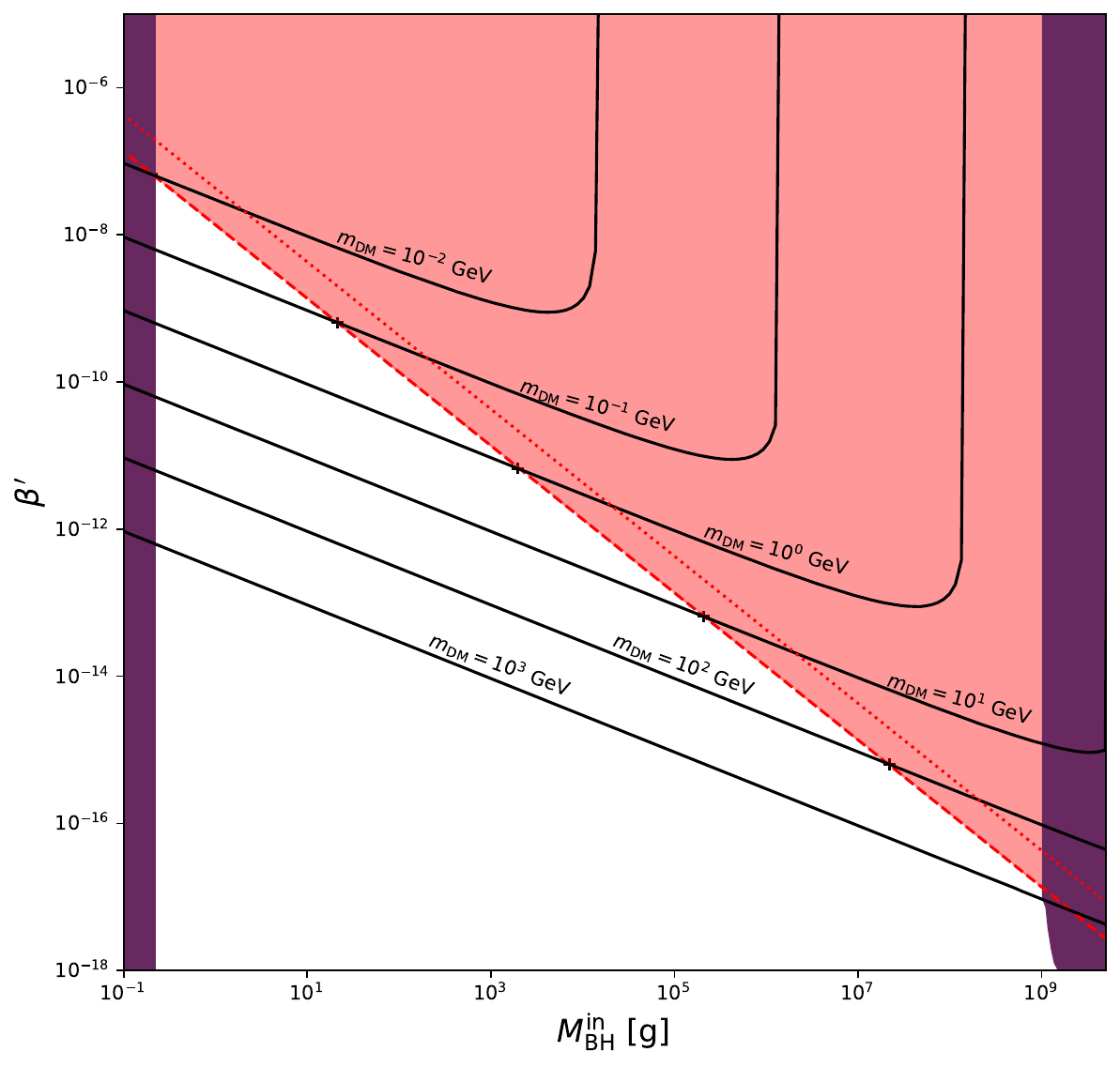}
 \caption{Figure showing the $\beta^\prime$ and $\MBHi$ values required to fully produce the correct relic abundance (solid black lines) for a range of dark matter masses. This is for the case where dark matter is a Fermion and all black holes are Schwarzschild. The black $+$ symbols indicate the precise point on the line where the transfer function, $T(k)$, becomes inconsistent with Lyman-$\alpha$ constraints, i.e. the $\alpha$ parameter of Eq.~\eqref{eq:Transfer_param} is $\geq 1.3\times10^{-2}\,{\rm Mpc}\,h^{-1}$. The red dashed line corresponds to the mean value found for $\eta$ as per Eq.~\eqref{eq:simple_betaeta}, where the values that violate the inequality are shaded red. The dotted line corresponds to the result reported in Ref.~\cite{Baldes:2020nuv}. The purple shaded regions signify the inflationary (low $\MBHi$)~\cite{Akrami:2018odb} and Big Bang Nucleosynthesis (high $\MBHi$) constraints on primordial black hole abundances~\cite{Carr:2009jm,Hasegawa:2019jsa,Carr:2020gox,Keith:2020jww}.}
 \label{fig:beta_mass_plot.pdf}
\end{figure}
Fig.~\ref{fig:beta_mass_plot.pdf} illustrates our procedure, in which we show the $\beta^\prime$ and $\MBHi$ values needed to produce the correct relic abundance through evaporation only for a range of $m_{\rm DM}$ values. Above these lines, dark matter would be overproduced and below it would be underproduced. Along each line, a black + symbol appears, which signifies the region of parameter space where the bound from Lyman-$\alpha$ is saturated, such that points with greater $\MBHi$ will be in tension with the observed matter power spectrum. 

In agreement with the literature before us, we find that for monochromatic distributions, the warm dark matter constraint, applied to dark matter with masses, $m_{\rm DM}\leq 10^3\,$ GeV, can be be well approximated by a simple relation, 
\begin{equation}
 \beta^\prime \leq \eta \left(\frac{\MPL}{\MBHi}\right)\,,
 \label{eq:simple_betaeta}
\end{equation}
where $\eta$ is a dimensionless parameter that varies depending on $a_{\star}$. In Refs.~\cite{Baldes:2020nuv,Auffinger:2020afu} the limit is presented as a multiple of $\beta_c$, the $\beta^\prime$ value required such that PBHs dominate at some point. Since there exists multiple definitions of $\beta_c$ in the literature~\cite{Gondolo:2020uqv, Baldes:2020nuv} and changes depending on $a_\star$, we believe that presenting our results in the form of Eq.~\eqref{eq:simple_betaeta} is easier to interpret. In order to determine a computational error on the value of $\eta$, we sample multiple $m_{\rm DM}$ values for a given $a_{\star}$ and calculate the standard error of the mean from the set of $\eta$ values.

Returning to Fig.~\ref{fig:beta_mass_plot.pdf}, we observe that the black + symbols, signifying where the warm dark matter constraints become relevant, align for different $m_{\rm DM}$ values such that one can write Eq.~\eqref{eq:simple_betaeta}. For Schwarzschild PBHs, fermion dark matter case shown in this figure, we find $\eta=(6.26\pm 0.08)\times 10^{-4}$,  this is the red dashed line in the figure, which borders the red shaded region representing where the full dark matter production through PBH evaporation is inconsistent with the matter power spectrum. This allows one to interpolate between the $m_{\rm DM}$ values shown in the figure. The dotted red line shows the warm dark matter constraint found in Refs~\cite{Baldes:2020nuv, Auffinger:2020afu}. This level of disagreement, we believe, is acceptable because of the potential for quite sizeable differences in the ratios of scale factors in Eq.~\eqref{eq:DMphasespace} and temperature in Eq.~\eqref{eq:tncdm}. \\

\begin{figure}
 \centering
 \includegraphics[width=\linewidth]{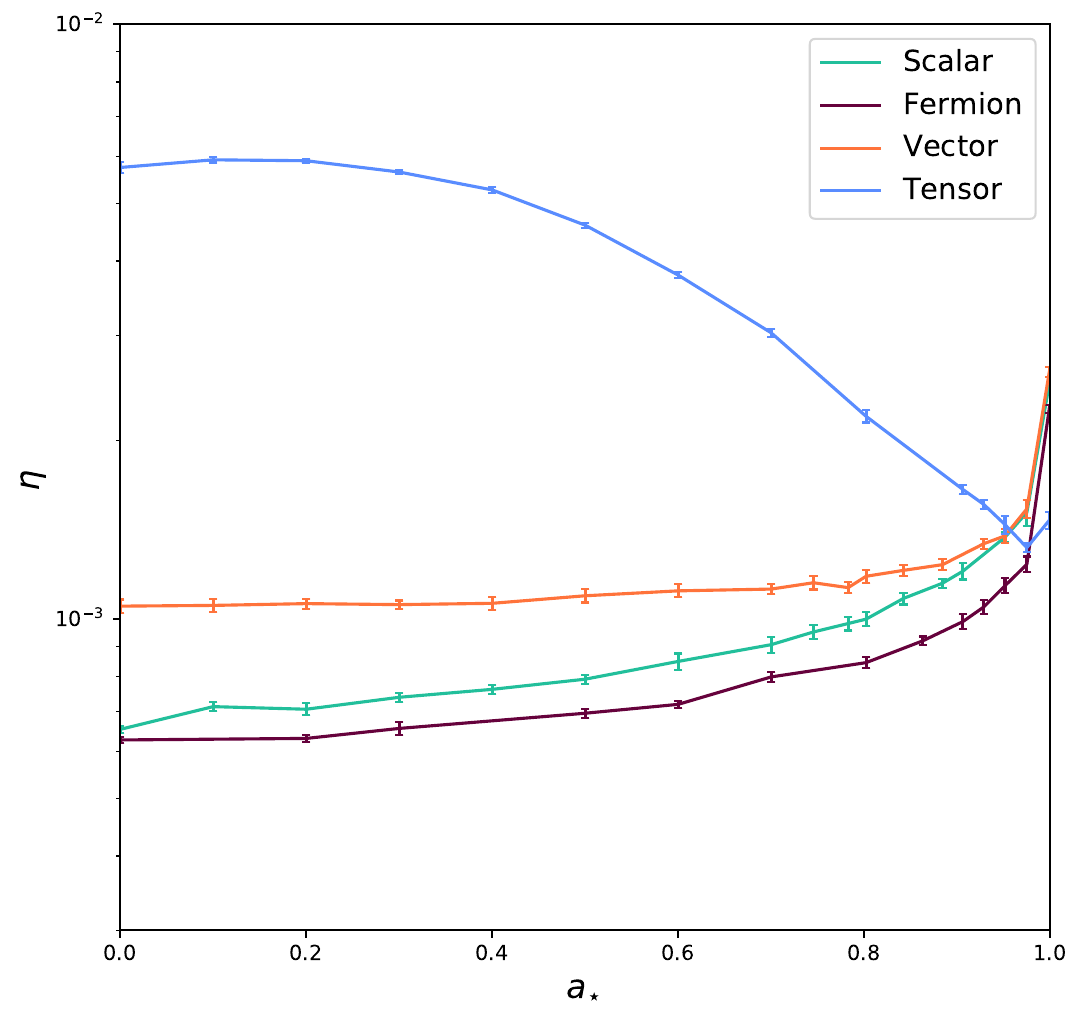}
 \caption{The variation of $\eta$ for different $a_{\star}$ values of the pbh distribution. Different colours correspond to different dark matter spins. The error bars indicate the computational variation of this value by way of the standard error of the mean calculated from the samples of different $m_{\rm DM}$ values. }
 \label{fig:vary_astar}
\end{figure}

In Fig.~\ref{fig:vary_astar}, we show our determination of $\eta$ and its standard error for scalar, fermion, vector and tensor dark matter particles with green, purple, orange and blue lines and error bars, respectively. One can see that there is very little variation of $\eta$ for each value of $a_\star$ thus validating the expression in Eq.~\eqref{eq:simple_betaeta}. However, as $a_\star$ varies, we see a substantial variation in $\eta$, where the effect is most pronounced for the tensor dark matter case. For the scalar, fermion and vector case, the warm dark matter constraint is hardly changing with $a_\star$ except for the very maximal case. This is qualitatively in agreement with Ref.~\cite{Masina:2021zpu}, where the author also concluded that changes in $a_{\star}$ mainly altered the warm dark matter constraints for the tensor particle. Our analysis, which solves both the Friedmann and Boltzmann equations, places this conclusion on firmer ground. 

When one compares the $a_\star=0$ case in Fig.\ref{fig:vary_astar}, we can make a comparison with warm dark matter constraints reported in Ref.~\cite{Auffinger:2020afu}. Similarly to the fermion case as mentioned above, we observe a slight deviation with our calculation. However the general pattern is consistent, as with Ref~\cite{Auffinger:2020afu} the scalar and fermion constraints are found to be very similar and the vector and tensor limits are less aggressive. More precisely the order of magnitude difference we see between scalar and tensor dark matter constraints is also reported in Ref.~\cite{Auffinger:2020afu}. This difference is largely attributable to the relative red-shift felt from particles with different intrinsic spin, as seen in section \ref{sec:deltaN}.

Throughout the analysis we additionally performed 2-dimensional fit for the parameterisation of Eq.(\ref{eq:Transfer_param}) were we fit for both $\alpha$ and $\mu$. We find in general quite good agreement with $\mu=1.12$, with only slight variations. Furthermore with the 1-dimensional fit, the $T(k)$ always matches well with simulation results. We leave a more sophisticated study for future work. This is particularly relevant for the situation where PBHs produce only a sub-component of dark matter, where Eq.(\ref{eq:Transfer_param}) is often not valid~\cite{Baur:2017stq, Auffinger:2020afu}. 

\section{Conclusions}\label{sec:concls}
In previous works, it was argued that the production of dark vector or tensor bosons (such as dark photons or hot gravitons) is enhanced for dark particles with spin larger than one when the evaporating black holes approach extremality. We have shown in this paper that this enhancement is mitigated when one thoroughly considers the dynamics of evaporation. While primordial black holes evaporate, their spin is generically dissipated first, whereas their mass is depleted in a second phase. Since the production of particles with spin larger than one is enhanced when primordial black holes are rotating, the largest contribution of the evaporation to dark radiation is emitted during the first phase of the evaporation. Consequently, these dark particles are subject to more significant redshifting, which can only be accounted for by solving the complete set of equations tracking the PBH's evolution and the Universe's.
On the contrary, the second phase of the evaporation dilutes this primary contribution, leading to a subsequent reduction of $\Delta N_{\rm eff}$. Considering this dilution, we showed that the maximum contribution of PBH evaporation to $\Delta N_{\rm eff}$ is reduced by a factor of $\mathcal O(10)$ for maximally rotating black holes. In the case of lower spin PBHs, in the limit $a_{\star}\to 0$, our results align with those in the literature.

Using {\tt FRISBHEE} we could also reconstruct the full phase-space distribution of light dark-matter particles when they are produced through PBH evaporation and use the code {\tt CLASS} to determine the corresponding matter power spectrum and consequently the warm dark matter constraints on dark matter produced from black hole evaporation. Our present study builds on previous works by introducing, for the first time, the fully evolved scale factor, $a(t)$ and plasma temperature $T$. We also provide a useful parametrization of warm dark matter limit for scalar, fermion, vector and tensor dark matter candidates, respectively. We show how this parametrization varies over a monochromatic Kerr PBH distribution of different $a_{\star}$ values.

Our results have been obtained assuming monochromatic distributions for mass and spin of the PBH. Thus, one may wonder how these conclusions would be affected if we consider more realistic spectra. We are currently investigating scenarios with extended distributions, and we plan to report our results in the near future.

\section*{Acknowledgments}
The authors would like to thank Julia Stadler and Deanna Hooper for their help with using {\tt CLASS}. 
AC is supported by the grant ``AstroCeNT: Particle Astrophysics Science and Technology Centre" carried out within the International Research Agendas programme of the Foundation for Polish Science financed by the European Union under the European Regional Development Fund. The work of LH is funded by the UK Science and Technology Facilities Council (STFC) under grant ST/P001246/1.
This work used the DiRAC@Durham facility managed by the Institute for Computational Cosmology on behalf of the STFC DiRAC HPC Facility (\href{www.dirac.ac.uk}{www.dirac.ac.uk}). The equipment was funded by BEIS capital funding via STFC capital grants ST/P002293/1, ST/R002371/1 and ST/S002502/1; Durham University and STFC operations grant ST/R000832/1. DiRAC is part of the National e-Infrastructure. 
This work has made use of the Hamilton HPC Service of Durham University.

\bibliographystyle{apsrev4-1}
\bibliography{PBH.bib}

\begin{thebibliography}{80}%
\makeatletter
\providecommand \@ifxundefined [1]{%
 \@ifx{#1\undefined}
}%
\providecommand \@ifnum [1]{%
 \ifnum #1\expandafter \@firstoftwo
 \else \expandafter \@secondoftwo
 \fi
}%
\providecommand \@ifx [1]{%
 \ifx #1\expandafter \@firstoftwo
 \else \expandafter \@secondoftwo
 \fi
}%
\providecommand \natexlab [1]{#1}%
\providecommand \enquote  [1]{``#1''}%
\providecommand \bibnamefont  [1]{#1}%
\providecommand \bibfnamefont [1]{#1}%
\providecommand \citenamefont [1]{#1}%
\providecommand \href@noop [0]{\@secondoftwo}%
\providecommand \href [0]{\begingroup \@sanitize@url \@href}%
\providecommand \@href[1]{\@@startlink{#1}\@@href}%
\providecommand \@@href[1]{\endgroup#1\@@endlink}%
\providecommand \@sanitize@url [0]{\catcode `\\12\catcode `\$12\catcode
  `\&12\catcode `\#12\catcode `\^12\catcode `\_12\catcode `\%12\relax}%
\providecommand \@@startlink[1]{}%
\providecommand \@@endlink[0]{}%
\providecommand \url  [0]{\begingroup\@sanitize@url \@url }%
\providecommand \@url [1]{\endgroup\@href {#1}{\urlprefix }}%
\providecommand \urlprefix  [0]{URL }%
\providecommand \Eprint [0]{\href }%
\providecommand \doibase [0]{http://dx.doi.org/}%
\providecommand \selectlanguage [0]{\@gobble}%
\providecommand \bibinfo  [0]{\@secondoftwo}%
\providecommand \bibfield  [0]{\@secondoftwo}%
\providecommand \translation [1]{[#1]}%
\providecommand \BibitemOpen [0]{}%
\providecommand \bibitemStop [0]{}%
\providecommand \bibitemNoStop [0]{.\EOS\space}%
\providecommand \EOS [0]{\spacefactor3000\relax}%
\providecommand \BibitemShut  [1]{\csname bibitem#1\endcsname}%
\let\auto@bib@innerbib\@empty
\bibitem [{\citenamefont {Abbott}\ \emph {et~al.}(2016)\citenamefont {Abbott}
  \emph {et~al.}}]{LIGOScientific:2016aoc}%
  \BibitemOpen
  \bibfield  {author} {\bibinfo {author} {\bibfnamefont {B.~P.}\ \bibnamefont
  {Abbott}} \emph {et~al.} (\bibinfo {collaboration} {LIGO Scientific,
  Virgo}),\ }\href {\doibase 10.1103/PhysRevLett.116.061102} {\bibfield
  {journal} {\bibinfo  {journal} {Phys. Rev. Lett.}\ }\textbf {\bibinfo
  {volume} {116}},\ \bibinfo {pages} {061102} (\bibinfo {year} {2016})},\
  \Eprint {http://arxiv.org/abs/1602.03837} {arXiv:1602.03837 [gr-qc]}
  \BibitemShut {NoStop}%
\bibitem [{\citenamefont {Abbott}\ \emph {et~al.}(2017)\citenamefont {Abbott}
  \emph {et~al.}}]{LIGOScientific:2017vwq}%
  \BibitemOpen
  \bibfield  {author} {\bibinfo {author} {\bibfnamefont {B.~P.}\ \bibnamefont
  {Abbott}} \emph {et~al.} (\bibinfo {collaboration} {LIGO Scientific,
  Virgo}),\ }\href {\doibase 10.1103/PhysRevLett.119.161101} {\bibfield
  {journal} {\bibinfo  {journal} {Phys. Rev. Lett.}\ }\textbf {\bibinfo
  {volume} {119}},\ \bibinfo {pages} {161101} (\bibinfo {year} {2017})},\
  \Eprint {http://arxiv.org/abs/1710.05832} {arXiv:1710.05832 [gr-qc]}
  \BibitemShut {NoStop}%
\bibitem [{\citenamefont {Auclair}\ \emph {et~al.}(2022)\citenamefont {Auclair}
  \emph {et~al.}}]{LISACosmologyWorkingGroup:2022jok}%
  \BibitemOpen
  \bibfield  {author} {\bibinfo {author} {\bibfnamefont {P.}~\bibnamefont
  {Auclair}} \emph {et~al.} (\bibinfo {collaboration} {LISA Cosmology Working
  Group}),\ }\href@noop {} {\  (\bibinfo {year} {2022})},\ \Eprint
  {http://arxiv.org/abs/2204.05434} {arXiv:2204.05434 [astro-ph.CO]}
  \BibitemShut {NoStop}%
\bibitem [{\citenamefont {Jeong}\ and\ \citenamefont
  {Takahashi}(2013)}]{Jeong:2013eza}%
  \BibitemOpen
  \bibfield  {author} {\bibinfo {author} {\bibfnamefont {K.~S.}\ \bibnamefont
  {Jeong}}\ and\ \bibinfo {author} {\bibfnamefont {F.}~\bibnamefont
  {Takahashi}},\ }\href {\doibase 10.1016/j.physletb.2013.07.001} {\bibfield
  {journal} {\bibinfo  {journal} {Phys. Lett. B}\ }\textbf {\bibinfo {volume}
  {725}},\ \bibinfo {pages} {134} (\bibinfo {year} {2013})},\ \Eprint
  {http://arxiv.org/abs/1305.6521} {arXiv:1305.6521 [hep-ph]} \BibitemShut
  {NoStop}%
\bibitem [{\citenamefont {Buen-Abad}\ \emph {et~al.}(2015)\citenamefont
  {Buen-Abad}, \citenamefont {Marques-Tavares},\ and\ \citenamefont
  {Schmaltz}}]{Buen-Abad:2015ova}%
  \BibitemOpen
  \bibfield  {author} {\bibinfo {author} {\bibfnamefont {M.~A.}\ \bibnamefont
  {Buen-Abad}}, \bibinfo {author} {\bibfnamefont {G.}~\bibnamefont
  {Marques-Tavares}}, \ and\ \bibinfo {author} {\bibfnamefont {M.}~\bibnamefont
  {Schmaltz}},\ }\href {\doibase 10.1103/PhysRevD.92.023531} {\bibfield
  {journal} {\bibinfo  {journal} {Phys. Rev. D}\ }\textbf {\bibinfo {volume}
  {92}},\ \bibinfo {pages} {023531} (\bibinfo {year} {2015})},\ \Eprint
  {http://arxiv.org/abs/1505.03542} {arXiv:1505.03542 [hep-ph]} \BibitemShut
  {NoStop}%
\bibitem [{\citenamefont {Lesgourgues}\ \emph {et~al.}(2016)\citenamefont
  {Lesgourgues}, \citenamefont {Marques-Tavares},\ and\ \citenamefont
  {Schmaltz}}]{Lesgourgues:2015wza}%
  \BibitemOpen
  \bibfield  {author} {\bibinfo {author} {\bibfnamefont {J.}~\bibnamefont
  {Lesgourgues}}, \bibinfo {author} {\bibfnamefont {G.}~\bibnamefont
  {Marques-Tavares}}, \ and\ \bibinfo {author} {\bibfnamefont {M.}~\bibnamefont
  {Schmaltz}},\ }\href {\doibase 10.1088/1475-7516/2016/02/037} {\bibfield
  {journal} {\bibinfo  {journal} {JCAP}\ }\textbf {\bibinfo {volume} {02}},\
  \bibinfo {pages} {037} (\bibinfo {year} {2016})},\ \Eprint
  {http://arxiv.org/abs/1507.04351} {arXiv:1507.04351 [astro-ph.CO]}
  \BibitemShut {NoStop}%
\bibitem [{\citenamefont {Buen-Abad}\ \emph {et~al.}(2018)\citenamefont
  {Buen-Abad}, \citenamefont {Schmaltz}, \citenamefont {Lesgourgues},\ and\
  \citenamefont {Brinckmann}}]{Buen-Abad:2017gxg}%
  \BibitemOpen
  \bibfield  {author} {\bibinfo {author} {\bibfnamefont {M.~A.}\ \bibnamefont
  {Buen-Abad}}, \bibinfo {author} {\bibfnamefont {M.}~\bibnamefont {Schmaltz}},
  \bibinfo {author} {\bibfnamefont {J.}~\bibnamefont {Lesgourgues}}, \ and\
  \bibinfo {author} {\bibfnamefont {T.}~\bibnamefont {Brinckmann}},\ }\href
  {\doibase 10.1088/1475-7516/2018/01/008} {\bibfield  {journal} {\bibinfo
  {journal} {JCAP}\ }\textbf {\bibinfo {volume} {01}},\ \bibinfo {pages} {008}
  (\bibinfo {year} {2018})},\ \Eprint {http://arxiv.org/abs/1708.09406}
  {arXiv:1708.09406 [astro-ph.CO]} \BibitemShut {NoStop}%
\bibitem [{\citenamefont {Blinov}\ and\ \citenamefont
  {Marques-Tavares}(2020)}]{Blinov:2020hmc}%
  \BibitemOpen
  \bibfield  {author} {\bibinfo {author} {\bibfnamefont {N.}~\bibnamefont
  {Blinov}}\ and\ \bibinfo {author} {\bibfnamefont {G.}~\bibnamefont
  {Marques-Tavares}},\ }\href {\doibase 10.1088/1475-7516/2020/09/029}
  {\bibfield  {journal} {\bibinfo  {journal} {JCAP}\ }\textbf {\bibinfo
  {volume} {09}},\ \bibinfo {pages} {029} (\bibinfo {year} {2020})},\ \Eprint
  {http://arxiv.org/abs/2003.08387} {arXiv:2003.08387 [astro-ph.CO]}
  \BibitemShut {NoStop}%
\bibitem [{\citenamefont {Chacko}\ \emph {et~al.}(2004)\citenamefont {Chacko},
  \citenamefont {Hall}, \citenamefont {Okui},\ and\ \citenamefont
  {Oliver}}]{Chacko:2003dt}%
  \BibitemOpen
  \bibfield  {author} {\bibinfo {author} {\bibfnamefont {Z.}~\bibnamefont
  {Chacko}}, \bibinfo {author} {\bibfnamefont {L.~J.}\ \bibnamefont {Hall}},
  \bibinfo {author} {\bibfnamefont {T.}~\bibnamefont {Okui}}, \ and\ \bibinfo
  {author} {\bibfnamefont {S.~J.}\ \bibnamefont {Oliver}},\ }\href {\doibase
  10.1103/PhysRevD.70.085008} {\bibfield  {journal} {\bibinfo  {journal} {Phys.
  Rev. D}\ }\textbf {\bibinfo {volume} {70}},\ \bibinfo {pages} {085008}
  (\bibinfo {year} {2004})},\ \Eprint {http://arxiv.org/abs/hep-ph/0312267}
  {arXiv:hep-ph/0312267} \BibitemShut {NoStop}%
\bibitem [{\citenamefont {Chacko}\ \emph {et~al.}(2005)\citenamefont {Chacko},
  \citenamefont {Hall}, \citenamefont {Oliver},\ and\ \citenamefont
  {Perelstein}}]{Chacko:2004cz}%
  \BibitemOpen
  \bibfield  {author} {\bibinfo {author} {\bibfnamefont {Z.}~\bibnamefont
  {Chacko}}, \bibinfo {author} {\bibfnamefont {L.~J.}\ \bibnamefont {Hall}},
  \bibinfo {author} {\bibfnamefont {S.~J.}\ \bibnamefont {Oliver}}, \ and\
  \bibinfo {author} {\bibfnamefont {M.}~\bibnamefont {Perelstein}},\ }\href
  {\doibase 10.1103/PhysRevLett.94.111801} {\bibfield  {journal} {\bibinfo
  {journal} {Phys. Rev. Lett.}\ }\textbf {\bibinfo {volume} {94}},\ \bibinfo
  {pages} {111801} (\bibinfo {year} {2005})},\ \Eprint
  {http://arxiv.org/abs/hep-ph/0405067} {arXiv:hep-ph/0405067} \BibitemShut
  {NoStop}%
\bibitem [{\citenamefont {Berlin}\ and\ \citenamefont
  {Blinov}(2018)}]{Berlin:2017ftj}%
  \BibitemOpen
  \bibfield  {author} {\bibinfo {author} {\bibfnamefont {A.}~\bibnamefont
  {Berlin}}\ and\ \bibinfo {author} {\bibfnamefont {N.}~\bibnamefont
  {Blinov}},\ }\href {\doibase 10.1103/PhysRevLett.120.021801} {\bibfield
  {journal} {\bibinfo  {journal} {Phys. Rev. Lett.}\ }\textbf {\bibinfo
  {volume} {120}},\ \bibinfo {pages} {021801} (\bibinfo {year} {2018})},\
  \Eprint {http://arxiv.org/abs/1706.07046} {arXiv:1706.07046 [hep-ph]}
  \BibitemShut {NoStop}%
\bibitem [{\citenamefont {Zel'dovich}\ and\ \citenamefont
  {Novikov}(1967)}]{Zeldovich:1967lct}%
  \BibitemOpen
  \bibfield  {author} {\bibinfo {author} {\bibfnamefont {Y.~B.}\ \bibnamefont
  {Zel'dovich}}\ and\ \bibinfo {author} {\bibfnamefont {I.~D.}\ \bibnamefont
  {Novikov}},\ }\href@noop {} {\bibfield  {journal} {\bibinfo  {journal}
  {Soviet Astron. AJ (Engl. Transl. ),}\ }\textbf {\bibinfo {volume} {10}},\
  \bibinfo {pages} {602} (\bibinfo {year} {1967})}\BibitemShut {NoStop}%
\bibitem [{\citenamefont {Hawking}(1971)}]{Hawking:1971ei}%
  \BibitemOpen
  \bibfield  {author} {\bibinfo {author} {\bibfnamefont {S.}~\bibnamefont
  {Hawking}},\ }\href@noop {} {\bibfield  {journal} {\bibinfo  {journal} {Mon.
  Not. Roy. Astron. Soc.}\ }\textbf {\bibinfo {volume} {152}},\ \bibinfo
  {pages} {75} (\bibinfo {year} {1971})}\BibitemShut {NoStop}%
\bibitem [{\citenamefont {Carr}\ and\ \citenamefont
  {Hawking}(1974)}]{Carr:1974nx}%
  \BibitemOpen
  \bibfield  {author} {\bibinfo {author} {\bibfnamefont {B.~J.}\ \bibnamefont
  {Carr}}\ and\ \bibinfo {author} {\bibfnamefont {S.~W.}\ \bibnamefont
  {Hawking}},\ }\href@noop {} {\bibfield  {journal} {\bibinfo  {journal} {Mon.
  Not. Roy. Astron. Soc.}\ }\textbf {\bibinfo {volume} {168}},\ \bibinfo
  {pages} {399} (\bibinfo {year} {1974})}\BibitemShut {NoStop}%
\bibitem [{\citenamefont {Inomata}\ \emph {et~al.}(2020)\citenamefont
  {Inomata}, \citenamefont {Kawasaki}, \citenamefont {Mukaida}, \citenamefont
  {Terada},\ and\ \citenamefont {Yanagida}}]{Inomata:2020lmk}%
  \BibitemOpen
  \bibfield  {author} {\bibinfo {author} {\bibfnamefont {K.}~\bibnamefont
  {Inomata}}, \bibinfo {author} {\bibfnamefont {M.}~\bibnamefont {Kawasaki}},
  \bibinfo {author} {\bibfnamefont {K.}~\bibnamefont {Mukaida}}, \bibinfo
  {author} {\bibfnamefont {T.}~\bibnamefont {Terada}}, \ and\ \bibinfo {author}
  {\bibfnamefont {T.~T.}\ \bibnamefont {Yanagida}},\ }\href {\doibase
  10.1103/PhysRevD.101.123533} {\bibfield  {journal} {\bibinfo  {journal}
  {Phys. Rev. D}\ }\textbf {\bibinfo {volume} {101}},\ \bibinfo {pages}
  {123533} (\bibinfo {year} {2020})},\ \Eprint
  {http://arxiv.org/abs/2003.10455} {arXiv:2003.10455 [astro-ph.CO]}
  \BibitemShut {NoStop}%
\bibitem [{\citenamefont {Dom\`enech}\ \emph {et~al.}(2021)\citenamefont
  {Dom\`enech}, \citenamefont {Lin},\ and\ \citenamefont
  {Sasaki}}]{Domenech:2020ssp}%
  \BibitemOpen
  \bibfield  {author} {\bibinfo {author} {\bibfnamefont {G.}~\bibnamefont
  {Dom\`enech}}, \bibinfo {author} {\bibfnamefont {C.}~\bibnamefont {Lin}}, \
  and\ \bibinfo {author} {\bibfnamefont {M.}~\bibnamefont {Sasaki}},\ }\href
  {\doibase 10.1088/1475-7516/2021/11/E01} {\bibfield  {journal} {\bibinfo
  {journal} {JCAP}\ }\textbf {\bibinfo {volume} {04}},\ \bibinfo {pages} {062}
  (\bibinfo {year} {2021})},\ \bibinfo {note} {[Erratum: JCAP 11, E01
  (2021)]},\ \Eprint {http://arxiv.org/abs/2012.08151} {arXiv:2012.08151
  [gr-qc]} \BibitemShut {NoStop}%
\bibitem [{\citenamefont {Hawking}(1974)}]{Hawking:1974rv}%
  \BibitemOpen
  \bibfield  {author} {\bibinfo {author} {\bibfnamefont {S.~W.}\ \bibnamefont
  {Hawking}},\ }\href {\doibase 10.1038/248030a0} {\bibfield  {journal}
  {\bibinfo  {journal} {Nature}\ }\textbf {\bibinfo {volume} {248}},\ \bibinfo
  {pages} {30} (\bibinfo {year} {1974})}\BibitemShut {NoStop}%
\bibitem [{\citenamefont {Hawking}(1975)}]{Hawking:1974sw}%
  \BibitemOpen
  \bibfield  {author} {\bibinfo {author} {\bibfnamefont {S.~W.}\ \bibnamefont
  {Hawking}},\ }\bibfield  {booktitle} {\emph {\bibinfo {booktitle} {{Euclidean
  quantum gravity}}},\ }\href {\doibase 10.1007/BF02345020, 10.1007/BF01608497}
  {\bibfield  {journal} {\bibinfo  {journal} {Commun. Math. Phys.}\ }\textbf
  {\bibinfo {volume} {43}},\ \bibinfo {pages} {199} (\bibinfo {year} {1975})},\
  \bibinfo {note} {[,167(1975)]}\BibitemShut {NoStop}%
\bibitem [{\citenamefont {Auffinger}(2022)}]{Auffinger:2022khh}%
  \BibitemOpen
  \bibfield  {author} {\bibinfo {author} {\bibfnamefont {J.}~\bibnamefont
  {Auffinger}},\ }\href@noop {} {\  (\bibinfo {year} {2022})},\ \Eprint
  {http://arxiv.org/abs/2206.02672} {arXiv:2206.02672 [astro-ph.CO]}
  \BibitemShut {NoStop}%
\bibitem [{\citenamefont {Dong}\ \emph {et~al.}(2016)\citenamefont {Dong},
  \citenamefont {Kinney},\ and\ \citenamefont {Stojkovic}}]{Dong:2015yjs}%
  \BibitemOpen
  \bibfield  {author} {\bibinfo {author} {\bibfnamefont {R.}~\bibnamefont
  {Dong}}, \bibinfo {author} {\bibfnamefont {W.~H.}\ \bibnamefont {Kinney}}, \
  and\ \bibinfo {author} {\bibfnamefont {D.}~\bibnamefont {Stojkovic}},\ }\href
  {\doibase 10.1088/1475-7516/2016/10/034} {\bibfield  {journal} {\bibinfo
  {journal} {JCAP}\ }\textbf {\bibinfo {volume} {10}},\ \bibinfo {pages} {034}
  (\bibinfo {year} {2016})},\ \Eprint {http://arxiv.org/abs/1511.05642}
  {arXiv:1511.05642 [astro-ph.CO]} \BibitemShut {NoStop}%
\bibitem [{\citenamefont {Page}(1976{\natexlab{a}})}]{Page:1976df}%
  \BibitemOpen
  \bibfield  {author} {\bibinfo {author} {\bibfnamefont {D.~N.}\ \bibnamefont
  {Page}},\ }\href {\doibase 10.1103/PhysRevD.13.198} {\bibfield  {journal}
  {\bibinfo  {journal} {Phys. Rev. D}\ }\textbf {\bibinfo {volume} {13}},\
  \bibinfo {pages} {198} (\bibinfo {year} {1976}{\natexlab{a}})}\BibitemShut
  {NoStop}%
\bibitem [{\citenamefont {Page}(1976{\natexlab{b}})}]{Page:1976ki}%
  \BibitemOpen
  \bibfield  {author} {\bibinfo {author} {\bibfnamefont {D.~N.}\ \bibnamefont
  {Page}},\ }\href {\doibase 10.1103/PhysRevD.14.3260} {\bibfield  {journal}
  {\bibinfo  {journal} {Phys. Rev. D}\ }\textbf {\bibinfo {volume} {14}},\
  \bibinfo {pages} {3260} (\bibinfo {year} {1976}{\natexlab{b}})}\BibitemShut
  {NoStop}%
\bibitem [{\citenamefont {Page}(1977)}]{Page:1977um}%
  \BibitemOpen
  \bibfield  {author} {\bibinfo {author} {\bibfnamefont {D.~N.}\ \bibnamefont
  {Page}},\ }\href {\doibase 10.1103/PhysRevD.16.2402} {\bibfield  {journal}
  {\bibinfo  {journal} {Phys. Rev. D}\ }\textbf {\bibinfo {volume} {16}},\
  \bibinfo {pages} {2402} (\bibinfo {year} {1977})}\BibitemShut {NoStop}%
\bibitem [{\citenamefont {Hooper}\ \emph {et~al.}(2019)\citenamefont {Hooper},
  \citenamefont {Krnjaic},\ and\ \citenamefont {McDermott}}]{Hooper:2019gtx}%
  \BibitemOpen
  \bibfield  {author} {\bibinfo {author} {\bibfnamefont {D.}~\bibnamefont
  {Hooper}}, \bibinfo {author} {\bibfnamefont {G.}~\bibnamefont {Krnjaic}}, \
  and\ \bibinfo {author} {\bibfnamefont {S.~D.}\ \bibnamefont {McDermott}},\
  }\href {\doibase 10.1007/JHEP08(2019)001} {\bibfield  {journal} {\bibinfo
  {journal} {JHEP}\ }\textbf {\bibinfo {volume} {08}},\ \bibinfo {pages} {001}
  (\bibinfo {year} {2019})},\ \Eprint {http://arxiv.org/abs/1905.01301}
  {arXiv:1905.01301 [hep-ph]} \BibitemShut {NoStop}%
\bibitem [{\citenamefont {Hooper}\ \emph {et~al.}(2020)\citenamefont {Hooper},
  \citenamefont {Krnjaic}, \citenamefont {March-Russell}, \citenamefont
  {McDermott},\ and\ \citenamefont {Petrossian-Byrne}}]{Hooper:2020evu}%
  \BibitemOpen
  \bibfield  {author} {\bibinfo {author} {\bibfnamefont {D.}~\bibnamefont
  {Hooper}}, \bibinfo {author} {\bibfnamefont {G.}~\bibnamefont {Krnjaic}},
  \bibinfo {author} {\bibfnamefont {J.}~\bibnamefont {March-Russell}}, \bibinfo
  {author} {\bibfnamefont {S.~D.}\ \bibnamefont {McDermott}}, \ and\ \bibinfo
  {author} {\bibfnamefont {R.}~\bibnamefont {Petrossian-Byrne}},\ }\href@noop
  {} {\  (\bibinfo {year} {2020})},\ \Eprint {http://arxiv.org/abs/2004.00618}
  {arXiv:2004.00618 [astro-ph.CO]} \BibitemShut {NoStop}%
\bibitem [{\citenamefont {Masina}(2020)}]{Masina:2020xhk}%
  \BibitemOpen
  \bibfield  {author} {\bibinfo {author} {\bibfnamefont {I.}~\bibnamefont
  {Masina}},\ }\href {\doibase 10.1140/epjp/s13360-020-00564-9} {\bibfield
  {journal} {\bibinfo  {journal} {Eur. Phys. J. Plus}\ }\textbf {\bibinfo
  {volume} {135}},\ \bibinfo {pages} {552} (\bibinfo {year} {2020})},\ \Eprint
  {http://arxiv.org/abs/2004.04740} {arXiv:2004.04740 [hep-ph]} \BibitemShut
  {NoStop}%
\bibitem [{\citenamefont {Masina}(2021)}]{Masina:2021zpu}%
  \BibitemOpen
  \bibfield  {author} {\bibinfo {author} {\bibfnamefont {I.}~\bibnamefont
  {Masina}},\ }\href {\doibase 10.1134/S0202289321040101} {\bibfield  {journal}
  {\bibinfo  {journal} {Grav. Cosmol.}\ }\textbf {\bibinfo {volume} {27}},\
  \bibinfo {pages} {315} (\bibinfo {year} {2021})},\ \Eprint
  {http://arxiv.org/abs/2103.13825} {arXiv:2103.13825 [gr-qc]} \BibitemShut
  {NoStop}%
\bibitem [{\citenamefont {Arbey}\ \emph {et~al.}(2021)\citenamefont {Arbey},
  \citenamefont {Auffinger}, \citenamefont {Sandick}, \citenamefont {Shams
  Es~Haghi},\ and\ \citenamefont {Sinha}}]{Arbey:2021ysg}%
  \BibitemOpen
  \bibfield  {author} {\bibinfo {author} {\bibfnamefont {A.}~\bibnamefont
  {Arbey}}, \bibinfo {author} {\bibfnamefont {J.}~\bibnamefont {Auffinger}},
  \bibinfo {author} {\bibfnamefont {P.}~\bibnamefont {Sandick}}, \bibinfo
  {author} {\bibfnamefont {B.}~\bibnamefont {Shams Es~Haghi}}, \ and\ \bibinfo
  {author} {\bibfnamefont {K.}~\bibnamefont {Sinha}},\ }\href@noop {} {\
  (\bibinfo {year} {2021})},\ \Eprint {http://arxiv.org/abs/2104.04051}
  {arXiv:2104.04051 [astro-ph.CO]} \BibitemShut {NoStop}%
\bibitem [{\citenamefont {Aiola}\ \emph {et~al.}(2022)\citenamefont {Aiola}
  \emph {et~al.}}]{CMB-HD:2022bsz}%
  \BibitemOpen
  \bibfield  {author} {\bibinfo {author} {\bibfnamefont {S.}~\bibnamefont
  {Aiola}} \emph {et~al.} (\bibinfo {collaboration} {CMB-HD}),\ }\href@noop {}
  {\  (\bibinfo {year} {2022})},\ \Eprint {http://arxiv.org/abs/2203.05728}
  {arXiv:2203.05728 [astro-ph.CO]} \BibitemShut {NoStop}%
\bibitem [{\citenamefont {Fujita}\ \emph {et~al.}(2014)\citenamefont {Fujita},
  \citenamefont {Kawasaki}, \citenamefont {Harigaya},\ and\ \citenamefont
  {Matsuda}}]{Fujita:2014hha}%
  \BibitemOpen
  \bibfield  {author} {\bibinfo {author} {\bibfnamefont {T.}~\bibnamefont
  {Fujita}}, \bibinfo {author} {\bibfnamefont {M.}~\bibnamefont {Kawasaki}},
  \bibinfo {author} {\bibfnamefont {K.}~\bibnamefont {Harigaya}}, \ and\
  \bibinfo {author} {\bibfnamefont {R.}~\bibnamefont {Matsuda}},\ }\href
  {\doibase 10.1103/PhysRevD.89.103501} {\bibfield  {journal} {\bibinfo
  {journal} {Phys. Rev.}\ }\textbf {\bibinfo {volume} {D89}},\ \bibinfo {pages}
  {103501} (\bibinfo {year} {2014})},\ \Eprint {http://arxiv.org/abs/1401.1909}
  {arXiv:1401.1909 [astro-ph.CO]} \BibitemShut {NoStop}%
\bibitem [{\citenamefont {Lennon}\ \emph {et~al.}(2018)\citenamefont {Lennon},
  \citenamefont {March-Russell}, \citenamefont {Petrossian-Byrne},\ and\
  \citenamefont {Tillim}}]{Lennon:2017tqq}%
  \BibitemOpen
  \bibfield  {author} {\bibinfo {author} {\bibfnamefont {O.}~\bibnamefont
  {Lennon}}, \bibinfo {author} {\bibfnamefont {J.}~\bibnamefont
  {March-Russell}}, \bibinfo {author} {\bibfnamefont {R.}~\bibnamefont
  {Petrossian-Byrne}}, \ and\ \bibinfo {author} {\bibfnamefont
  {H.}~\bibnamefont {Tillim}},\ }\href {\doibase 10.1088/1475-7516/2018/04/009}
  {\bibfield  {journal} {\bibinfo  {journal} {JCAP}\ }\textbf {\bibinfo
  {volume} {1804}},\ \bibinfo {pages} {009} (\bibinfo {year} {2018})},\ \Eprint
  {http://arxiv.org/abs/1712.07664} {arXiv:1712.07664 [hep-ph]} \BibitemShut
  {NoStop}%
\bibitem [{\citenamefont {Morrison}\ \emph {et~al.}(2019)\citenamefont
  {Morrison}, \citenamefont {Profumo},\ and\ \citenamefont
  {Yu}}]{Morrison:2018xla}%
  \BibitemOpen
  \bibfield  {author} {\bibinfo {author} {\bibfnamefont {L.}~\bibnamefont
  {Morrison}}, \bibinfo {author} {\bibfnamefont {S.}~\bibnamefont {Profumo}}, \
  and\ \bibinfo {author} {\bibfnamefont {Y.}~\bibnamefont {Yu}},\ }\href
  {\doibase 10.1088/1475-7516/2019/05/005} {\bibfield  {journal} {\bibinfo
  {journal} {JCAP}\ }\textbf {\bibinfo {volume} {1905}},\ \bibinfo {pages}
  {005} (\bibinfo {year} {2019})},\ \Eprint {http://arxiv.org/abs/1812.10606}
  {arXiv:1812.10606 [astro-ph.CO]} \BibitemShut {NoStop}%
\bibitem [{\citenamefont {Baldes}\ \emph {et~al.}(2020)\citenamefont {Baldes},
  \citenamefont {Decant}, \citenamefont {Hooper},\ and\ \citenamefont
  {Lopez-Honorez}}]{Baldes:2020nuv}%
  \BibitemOpen
  \bibfield  {author} {\bibinfo {author} {\bibfnamefont {I.}~\bibnamefont
  {Baldes}}, \bibinfo {author} {\bibfnamefont {Q.}~\bibnamefont {Decant}},
  \bibinfo {author} {\bibfnamefont {D.~C.}\ \bibnamefont {Hooper}}, \ and\
  \bibinfo {author} {\bibfnamefont {L.}~\bibnamefont {Lopez-Honorez}},\ }\href
  {\doibase 10.1088/1475-7516/2020/08/045} {\bibfield  {journal} {\bibinfo
  {journal} {JCAP}\ }\textbf {\bibinfo {volume} {08}},\ \bibinfo {pages} {045}
  (\bibinfo {year} {2020})},\ \Eprint {http://arxiv.org/abs/2004.14773}
  {arXiv:2004.14773 [astro-ph.CO]} \BibitemShut {NoStop}%
\bibitem [{\citenamefont {Gondolo}\ \emph {et~al.}(2020)\citenamefont
  {Gondolo}, \citenamefont {Sandick},\ and\ \citenamefont {Shams
  Es~Haghi}}]{Gondolo:2020uqv}%
  \BibitemOpen
  \bibfield  {author} {\bibinfo {author} {\bibfnamefont {P.}~\bibnamefont
  {Gondolo}}, \bibinfo {author} {\bibfnamefont {P.}~\bibnamefont {Sandick}}, \
  and\ \bibinfo {author} {\bibfnamefont {B.}~\bibnamefont {Shams Es~Haghi}},\
  }\href {\doibase 10.1103/PhysRevD.102.095018} {\bibfield  {journal} {\bibinfo
   {journal} {Phys. Rev. D}\ }\textbf {\bibinfo {volume} {102}},\ \bibinfo
  {pages} {095018} (\bibinfo {year} {2020})},\ \Eprint
  {http://arxiv.org/abs/2009.02424} {arXiv:2009.02424 [hep-ph]} \BibitemShut
  {NoStop}%
\bibitem [{\citenamefont {Bernal}\ and\ \citenamefont
  {Zapata}(2021{\natexlab{a}})}]{Bernal:2020kse}%
  \BibitemOpen
  \bibfield  {author} {\bibinfo {author} {\bibfnamefont {N.}~\bibnamefont
  {Bernal}}\ and\ \bibinfo {author} {\bibfnamefont {O.}~\bibnamefont
  {Zapata}},\ }\href {\doibase 10.1088/1475-7516/2021/03/007} {\bibfield
  {journal} {\bibinfo  {journal} {JCAP}\ }\textbf {\bibinfo {volume} {03}},\
  \bibinfo {pages} {007} (\bibinfo {year} {2021}{\natexlab{a}})},\ \Eprint
  {http://arxiv.org/abs/2010.09725} {arXiv:2010.09725 [hep-ph]} \BibitemShut
  {NoStop}%
\bibitem [{\citenamefont {Bernal}\ and\ \citenamefont
  {Zapata}(2021{\natexlab{b}})}]{Bernal:2020ili}%
  \BibitemOpen
  \bibfield  {author} {\bibinfo {author} {\bibfnamefont {N.}~\bibnamefont
  {Bernal}}\ and\ \bibinfo {author} {\bibfnamefont {O.}~\bibnamefont
  {Zapata}},\ }\href {\doibase 10.1016/j.physletb.2021.136129} {\bibfield
  {journal} {\bibinfo  {journal} {Phys. Lett. B}\ }\textbf {\bibinfo {volume}
  {815}},\ \bibinfo {pages} {136129} (\bibinfo {year} {2021}{\natexlab{b}})},\
  \Eprint {http://arxiv.org/abs/2011.02510} {arXiv:2011.02510 [hep-ph]}
  \BibitemShut {NoStop}%
\bibitem [{\citenamefont {Auffinger}\ \emph {et~al.}(2021)\citenamefont
  {Auffinger}, \citenamefont {Masina},\ and\ \citenamefont
  {Orlando}}]{Auffinger:2020afu}%
  \BibitemOpen
  \bibfield  {author} {\bibinfo {author} {\bibfnamefont {J.}~\bibnamefont
  {Auffinger}}, \bibinfo {author} {\bibfnamefont {I.}~\bibnamefont {Masina}}, \
  and\ \bibinfo {author} {\bibfnamefont {G.}~\bibnamefont {Orlando}},\ }\href
  {\doibase 10.1140/epjp/s13360-021-01247-9} {\bibfield  {journal} {\bibinfo
  {journal} {Eur. Phys. J. Plus}\ }\textbf {\bibinfo {volume} {136}},\ \bibinfo
  {pages} {261} (\bibinfo {year} {2021})},\ \Eprint
  {http://arxiv.org/abs/2012.09867} {arXiv:2012.09867 [hep-ph]} \BibitemShut
  {NoStop}%
\bibitem [{\citenamefont {Bernal}\ and\ \citenamefont
  {Zapata}(2021{\natexlab{c}})}]{Bernal:2020bjf}%
  \BibitemOpen
  \bibfield  {author} {\bibinfo {author} {\bibfnamefont {N.}~\bibnamefont
  {Bernal}}\ and\ \bibinfo {author} {\bibfnamefont {O.}~\bibnamefont
  {Zapata}},\ }\href {\doibase 10.1088/1475-7516/2021/03/015} {\bibfield
  {journal} {\bibinfo  {journal} {JCAP}\ }\textbf {\bibinfo {volume} {03}},\
  \bibinfo {pages} {015} (\bibinfo {year} {2021}{\natexlab{c}})},\ \Eprint
  {http://arxiv.org/abs/2011.12306} {arXiv:2011.12306 [astro-ph.CO]}
  \BibitemShut {NoStop}%
\bibitem [{\citenamefont {Jyoti~Das}\ \emph {et~al.}(2021)\citenamefont
  {Jyoti~Das}, \citenamefont {Mahanta},\ and\ \citenamefont
  {Borah}}]{JyotiDas:2021shi}%
  \BibitemOpen
  \bibfield  {author} {\bibinfo {author} {\bibfnamefont {S.}~\bibnamefont
  {Jyoti~Das}}, \bibinfo {author} {\bibfnamefont {D.}~\bibnamefont {Mahanta}},
  \ and\ \bibinfo {author} {\bibfnamefont {D.}~\bibnamefont {Borah}},\
  }\href@noop {} {\  (\bibinfo {year} {2021})},\ \Eprint
  {http://arxiv.org/abs/2104.14496} {arXiv:2104.14496 [hep-ph]} \BibitemShut
  {NoStop}%
\bibitem [{\citenamefont {Cheek}\ \emph
  {et~al.}(2022{\natexlab{a}})\citenamefont {Cheek}, \citenamefont {Heurtier},
  \citenamefont {Perez-Gonzalez},\ and\ \citenamefont
  {Turner}}]{Cheek:2021odj}%
  \BibitemOpen
  \bibfield  {author} {\bibinfo {author} {\bibfnamefont {A.}~\bibnamefont
  {Cheek}}, \bibinfo {author} {\bibfnamefont {L.}~\bibnamefont {Heurtier}},
  \bibinfo {author} {\bibfnamefont {Y.~F.}\ \bibnamefont {Perez-Gonzalez}}, \
  and\ \bibinfo {author} {\bibfnamefont {J.}~\bibnamefont {Turner}},\ }\href
  {\doibase 10.1103/PhysRevD.105.015022} {\bibfield  {journal} {\bibinfo
  {journal} {Phys. Rev. D}\ }\textbf {\bibinfo {volume} {105}},\ \bibinfo
  {pages} {015022} (\bibinfo {year} {2022}{\natexlab{a}})},\ \Eprint
  {http://arxiv.org/abs/2107.00013} {arXiv:2107.00013 [hep-ph]} \BibitemShut
  {NoStop}%
\bibitem [{\citenamefont {Cheek}\ \emph
  {et~al.}(2022{\natexlab{b}})\citenamefont {Cheek}, \citenamefont {Heurtier},
  \citenamefont {Perez-Gonzalez},\ and\ \citenamefont
  {Turner}}]{Cheek:2021cfe}%
  \BibitemOpen
  \bibfield  {author} {\bibinfo {author} {\bibfnamefont {A.}~\bibnamefont
  {Cheek}}, \bibinfo {author} {\bibfnamefont {L.}~\bibnamefont {Heurtier}},
  \bibinfo {author} {\bibfnamefont {Y.~F.}\ \bibnamefont {Perez-Gonzalez}}, \
  and\ \bibinfo {author} {\bibfnamefont {J.}~\bibnamefont {Turner}},\ }\href
  {\doibase 10.1103/PhysRevD.105.015023} {\bibfield  {journal} {\bibinfo
  {journal} {Phys. Rev. D}\ }\textbf {\bibinfo {volume} {105}},\ \bibinfo
  {pages} {015023} (\bibinfo {year} {2022}{\natexlab{b}})},\ \Eprint
  {http://arxiv.org/abs/2107.00016} {arXiv:2107.00016 [hep-ph]} \BibitemShut
  {NoStop}%
\bibitem [{\citenamefont {Sandick}\ \emph {et~al.}(2021)\citenamefont
  {Sandick}, \citenamefont {Es~Haghi},\ and\ \citenamefont
  {Sinha}}]{Sandick:2021gew}%
  \BibitemOpen
  \bibfield  {author} {\bibinfo {author} {\bibfnamefont {P.}~\bibnamefont
  {Sandick}}, \bibinfo {author} {\bibfnamefont {B.~S.}\ \bibnamefont
  {Es~Haghi}}, \ and\ \bibinfo {author} {\bibfnamefont {K.}~\bibnamefont
  {Sinha}},\ }\href {\doibase 10.1103/PhysRevD.104.083523} {\bibfield
  {journal} {\bibinfo  {journal} {Phys. Rev. D}\ }\textbf {\bibinfo {volume}
  {104}},\ \bibinfo {pages} {083523} (\bibinfo {year} {2021})},\ \Eprint
  {http://arxiv.org/abs/2108.08329} {arXiv:2108.08329 [astro-ph.CO]}
  \BibitemShut {NoStop}%
\bibitem [{\citenamefont {Bernal}\ \emph {et~al.}(2022)\citenamefont {Bernal},
  \citenamefont {Perez-Gonzalez},\ and\ \citenamefont {Xu}}]{Bernal:2022oha}%
  \BibitemOpen
  \bibfield  {author} {\bibinfo {author} {\bibfnamefont {N.}~\bibnamefont
  {Bernal}}, \bibinfo {author} {\bibfnamefont {Y.~F.}\ \bibnamefont
  {Perez-Gonzalez}}, \ and\ \bibinfo {author} {\bibfnamefont {Y.}~\bibnamefont
  {Xu}},\ }\href@noop {} {\  (\bibinfo {year} {2022})},\ \Eprint
  {http://arxiv.org/abs/2205.11522} {arXiv:2205.11522 [hep-ph]} \BibitemShut
  {NoStop}%
\bibitem [{\citenamefont {{Lesgourgues}}(2011)}]{CLASSI}%
  \BibitemOpen
  \bibfield  {author} {\bibinfo {author} {\bibfnamefont {J.}~\bibnamefont
  {{Lesgourgues}}},\ }\href@noop {} {\bibfield  {journal} {\bibinfo  {journal}
  {arXiv e-prints}\ ,\ \bibinfo {eid} {arXiv:1104.2932}} (\bibinfo {year}
  {2011})},\ \Eprint {http://arxiv.org/abs/1104.2932} {arXiv:1104.2932
  [astro-ph.IM]} \BibitemShut {NoStop}%
\bibitem [{\citenamefont {{Blas}}\ \emph {et~al.}(2011)\citenamefont {{Blas}},
  \citenamefont {{Lesgourgues}},\ and\ \citenamefont {{Tram}}}]{CLASSII}%
  \BibitemOpen
  \bibfield  {author} {\bibinfo {author} {\bibfnamefont {D.}~\bibnamefont
  {{Blas}}}, \bibinfo {author} {\bibfnamefont {J.}~\bibnamefont
  {{Lesgourgues}}}, \ and\ \bibinfo {author} {\bibfnamefont {T.}~\bibnamefont
  {{Tram}}},\ }\href {\doibase 10.1088/1475-7516/2011/07/034} {\bibfield
  {journal} {\bibinfo  {journal} {\it{JCAP}}\ }\textbf {\bibinfo {volume}
  {2011}},\ \bibinfo {eid} {034} (\bibinfo {year} {2011})},\ \Eprint
  {http://arxiv.org/abs/1104.2933} {arXiv:1104.2933 [astro-ph.CO]} \BibitemShut
  {NoStop}%
\bibitem [{\citenamefont {{Lesgourgues}}\ and\ \citenamefont
  {{Tram}}(2011)}]{CLASSIV}%
  \BibitemOpen
  \bibfield  {author} {\bibinfo {author} {\bibfnamefont {J.}~\bibnamefont
  {{Lesgourgues}}}\ and\ \bibinfo {author} {\bibfnamefont {T.}~\bibnamefont
  {{Tram}}},\ }\href {\doibase 10.1088/1475-7516/2011/09/032} {\bibfield
  {journal} {\bibinfo  {journal} {\it{JCAP}}\ }\textbf {\bibinfo {volume}
  {2011}},\ \bibinfo {eid} {032} (\bibinfo {year} {2011})},\ \Eprint
  {http://arxiv.org/abs/1104.2935} {arXiv:1104.2935 [astro-ph.CO]} \BibitemShut
  {NoStop}%
\bibitem [{\citenamefont {Chandrasekhar}\ and\ \citenamefont
  {Detweiler}(1975)}]{Chandrasekhar:1975zz}%
  \BibitemOpen
  \bibfield  {author} {\bibinfo {author} {\bibfnamefont {S.}~\bibnamefont
  {Chandrasekhar}}\ and\ \bibinfo {author} {\bibfnamefont {S.~L.}\ \bibnamefont
  {Detweiler}},\ }\href {\doibase 10.1098/rspa.1975.0130} {\bibfield  {journal}
  {\bibinfo  {journal} {Proc. Roy. Soc. Lond. A}\ }\textbf {\bibinfo {volume}
  {345}},\ \bibinfo {pages} {145} (\bibinfo {year} {1975})}\BibitemShut
  {NoStop}%
\bibitem [{\citenamefont {Chandrasekhar}\ and\ \citenamefont
  {Detweiler}(1976)}]{Chandrasekhar:1976zz}%
  \BibitemOpen
  \bibfield  {author} {\bibinfo {author} {\bibfnamefont {S.}~\bibnamefont
  {Chandrasekhar}}\ and\ \bibinfo {author} {\bibfnamefont {S.~L.}\ \bibnamefont
  {Detweiler}},\ }\href {\doibase 10.1098/rspa.1976.0101} {\bibfield  {journal}
  {\bibinfo  {journal} {Proc. Roy. Soc. Lond. A}\ }\textbf {\bibinfo {volume}
  {350}},\ \bibinfo {pages} {165} (\bibinfo {year} {1976})}\BibitemShut
  {NoStop}%
\bibitem [{\citenamefont {Chandrasekhar}\ and\ \citenamefont
  {Detweiler}(1977)}]{Chandrasekhar:1977kf}%
  \BibitemOpen
  \bibfield  {author} {\bibinfo {author} {\bibfnamefont {S.}~\bibnamefont
  {Chandrasekhar}}\ and\ \bibinfo {author} {\bibfnamefont {S.~L.}\ \bibnamefont
  {Detweiler}},\ }\href {\doibase 10.1098/rspa.1977.0002} {\bibfield  {journal}
  {\bibinfo  {journal} {Proc. Roy. Soc. Lond. A}\ }\textbf {\bibinfo {volume}
  {352}},\ \bibinfo {pages} {325} (\bibinfo {year} {1977})}\BibitemShut
  {NoStop}%
\bibitem [{\citenamefont {MacGibbon}\ and\ \citenamefont
  {Webber}(1990)}]{PhysRevD.41.3052}%
  \BibitemOpen
  \bibfield  {author} {\bibinfo {author} {\bibfnamefont {J.~H.}\ \bibnamefont
  {MacGibbon}}\ and\ \bibinfo {author} {\bibfnamefont {B.~R.}\ \bibnamefont
  {Webber}},\ }\href {\doibase 10.1103/PhysRevD.41.3052} {\bibfield  {journal}
  {\bibinfo  {journal} {Phys. Rev. D}\ }\textbf {\bibinfo {volume} {41}},\
  \bibinfo {pages} {3052} (\bibinfo {year} {1990})}\BibitemShut {NoStop}%
\bibitem [{\citenamefont {MacGibbon}(1991)}]{PhysRevD.44.376}%
  \BibitemOpen
  \bibfield  {author} {\bibinfo {author} {\bibfnamefont {J.~H.}\ \bibnamefont
  {MacGibbon}},\ }\href {\doibase 10.1103/PhysRevD.44.376} {\bibfield
  {journal} {\bibinfo  {journal} {Phys. Rev. D}\ }\textbf {\bibinfo {volume}
  {44}},\ \bibinfo {pages} {376} (\bibinfo {year} {1991})}\BibitemShut
  {NoStop}%
\bibitem [{\citenamefont {Fishbach}\ \emph {et~al.}(2017)\citenamefont
  {Fishbach}, \citenamefont {Holz},\ and\ \citenamefont
  {Farr}}]{Fishbach:2017dwv}%
  \BibitemOpen
  \bibfield  {author} {\bibinfo {author} {\bibfnamefont {M.}~\bibnamefont
  {Fishbach}}, \bibinfo {author} {\bibfnamefont {D.~E.}\ \bibnamefont {Holz}},
  \ and\ \bibinfo {author} {\bibfnamefont {B.}~\bibnamefont {Farr}},\ }\href
  {\doibase 10.3847/2041-8213/aa7045} {\bibfield  {journal} {\bibinfo
  {journal} {Astrophys. J. Lett.}\ }\textbf {\bibinfo {volume} {840}},\
  \bibinfo {pages} {L24} (\bibinfo {year} {2017})},\ \Eprint
  {http://arxiv.org/abs/1703.06869} {arXiv:1703.06869 [astro-ph.HE]}
  \BibitemShut {NoStop}%
\bibitem [{\citenamefont {Mangano}\ \emph {et~al.}(2002)\citenamefont
  {Mangano}, \citenamefont {Miele}, \citenamefont {Pastor},\ and\ \citenamefont
  {Peloso}}]{Mangano:2001iu}%
  \BibitemOpen
  \bibfield  {author} {\bibinfo {author} {\bibfnamefont {G.}~\bibnamefont
  {Mangano}}, \bibinfo {author} {\bibfnamefont {G.}~\bibnamefont {Miele}},
  \bibinfo {author} {\bibfnamefont {S.}~\bibnamefont {Pastor}}, \ and\ \bibinfo
  {author} {\bibfnamefont {M.}~\bibnamefont {Peloso}},\ }\href {\doibase
  10.1016/S0370-2693(02)01622-2} {\bibfield  {journal} {\bibinfo  {journal}
  {Phys. Lett. B}\ }\textbf {\bibinfo {volume} {534}},\ \bibinfo {pages} {8}
  (\bibinfo {year} {2002})},\ \Eprint {http://arxiv.org/abs/astro-ph/0111408}
  {arXiv:astro-ph/0111408} \BibitemShut {NoStop}%
\bibitem [{\citenamefont {de~Salas}\ and\ \citenamefont
  {Pastor}(2016)}]{deSalas:2016ztq}%
  \BibitemOpen
  \bibfield  {author} {\bibinfo {author} {\bibfnamefont {P.~F.}\ \bibnamefont
  {de~Salas}}\ and\ \bibinfo {author} {\bibfnamefont {S.}~\bibnamefont
  {Pastor}},\ }\href {\doibase 10.1088/1475-7516/2016/07/051} {\bibfield
  {journal} {\bibinfo  {journal} {JCAP}\ }\textbf {\bibinfo {volume} {07}},\
  \bibinfo {pages} {051} (\bibinfo {year} {2016})},\ \Eprint
  {http://arxiv.org/abs/1606.06986} {arXiv:1606.06986 [hep-ph]} \BibitemShut
  {NoStop}%
\bibitem [{\citenamefont {Carr}\ \emph {et~al.}(2020)\citenamefont {Carr},
  \citenamefont {Kohri}, \citenamefont {Sendouda},\ and\ \citenamefont
  {Yokoyama}}]{Carr:2020gox}%
  \BibitemOpen
  \bibfield  {author} {\bibinfo {author} {\bibfnamefont {B.}~\bibnamefont
  {Carr}}, \bibinfo {author} {\bibfnamefont {K.}~\bibnamefont {Kohri}},
  \bibinfo {author} {\bibfnamefont {Y.}~\bibnamefont {Sendouda}}, \ and\
  \bibinfo {author} {\bibfnamefont {J.}~\bibnamefont {Yokoyama}},\ }\href@noop
  {} {\  (\bibinfo {year} {2020})},\ \Eprint {http://arxiv.org/abs/2002.12778}
  {arXiv:2002.12778 [astro-ph.CO]} \BibitemShut {NoStop}%
\bibitem [{\citenamefont {Guti\'errez}\ \emph {et~al.}(2018)\citenamefont
  {Guti\'errez}, \citenamefont {Vieyro},\ and\ \citenamefont
  {Romero}}]{Gutierrez:2017ibk}%
  \BibitemOpen
  \bibfield  {author} {\bibinfo {author} {\bibfnamefont {E.~M.}\ \bibnamefont
  {Guti\'errez}}, \bibinfo {author} {\bibfnamefont {F.~L.}\ \bibnamefont
  {Vieyro}}, \ and\ \bibinfo {author} {\bibfnamefont {G.~E.}\ \bibnamefont
  {Romero}},\ }\href {\doibase 10.1093/mnras/stx2654} {\bibfield  {journal}
  {\bibinfo  {journal} {Mon. Not. Roy. Astron. Soc.}\ }\textbf {\bibinfo
  {volume} {473}},\ \bibinfo {pages} {5385} (\bibinfo {year} {2018})},\ \Eprint
  {http://arxiv.org/abs/1710.03061} {arXiv:1710.03061 [astro-ph.CO]}
  \BibitemShut {NoStop}%
\bibitem [{\citenamefont {Aghanim}\ \emph {et~al.}(2020)\citenamefont {Aghanim}
  \emph {et~al.}}]{Planck:2018vyg}%
  \BibitemOpen
  \bibfield  {author} {\bibinfo {author} {\bibfnamefont {N.}~\bibnamefont
  {Aghanim}} \emph {et~al.} (\bibinfo {collaboration} {Planck}),\ }\href
  {\doibase 10.1051/0004-6361/201833910} {\bibfield  {journal} {\bibinfo
  {journal} {Astron. Astrophys.}\ }\textbf {\bibinfo {volume} {641}},\ \bibinfo
  {pages} {A6} (\bibinfo {year} {2020})},\ \bibinfo {note} {[Erratum:
  Astron.Astrophys. 652, C4 (2021)]},\ \Eprint
  {http://arxiv.org/abs/1807.06209} {arXiv:1807.06209 [astro-ph.CO]}
  \BibitemShut {NoStop}%
\bibitem [{\citenamefont {Di~Valentino}\ \emph {et~al.}(2018)\citenamefont
  {Di~Valentino} \emph {et~al.}}]{CORE:2016npo}%
  \BibitemOpen
  \bibfield  {author} {\bibinfo {author} {\bibfnamefont {E.}~\bibnamefont
  {Di~Valentino}} \emph {et~al.} (\bibinfo {collaboration} {CORE}),\ }\href
  {\doibase 10.1088/1475-7516/2018/04/017} {\bibfield  {journal} {\bibinfo
  {journal} {JCAP}\ }\textbf {\bibinfo {volume} {04}},\ \bibinfo {pages} {017}
  (\bibinfo {year} {2018})},\ \Eprint {http://arxiv.org/abs/1612.00021}
  {arXiv:1612.00021 [astro-ph.CO]} \BibitemShut {NoStop}%
\bibitem [{\citenamefont {Aghamousa}\ \emph {et~al.}(2016)\citenamefont
  {Aghamousa} \emph {et~al.}}]{DESI:2016fyo}%
  \BibitemOpen
  \bibfield  {author} {\bibinfo {author} {\bibfnamefont {A.}~\bibnamefont
  {Aghamousa}} \emph {et~al.} (\bibinfo {collaboration} {DESI}),\ }\href@noop
  {} {\  (\bibinfo {year} {2016})},\ \Eprint {http://arxiv.org/abs/1611.00036}
  {arXiv:1611.00036 [astro-ph.IM]} \BibitemShut {NoStop}%
\bibitem [{\citenamefont {Laureijs}\ \emph {et~al.}(2011)\citenamefont
  {Laureijs} \emph {et~al.}}]{EUCLID:2011zbd}%
  \BibitemOpen
  \bibfield  {author} {\bibinfo {author} {\bibfnamefont {R.}~\bibnamefont
  {Laureijs}} \emph {et~al.} (\bibinfo {collaboration} {EUCLID}),\ }\href@noop
  {} {\  (\bibinfo {year} {2011})},\ \Eprint {http://arxiv.org/abs/1110.3193}
  {arXiv:1110.3193 [astro-ph.CO]} \BibitemShut {NoStop}%
\bibitem [{\citenamefont {Lattanzi}\ and\ \citenamefont
  {Gerbino}(2018)}]{Lattanzi:2017ubx}%
  \BibitemOpen
  \bibfield  {author} {\bibinfo {author} {\bibfnamefont {M.}~\bibnamefont
  {Lattanzi}}\ and\ \bibinfo {author} {\bibfnamefont {M.}~\bibnamefont
  {Gerbino}},\ }\href {\doibase 10.3389/fphy.2017.00070} {\bibfield  {journal}
  {\bibinfo  {journal} {Front. in Phys.}\ }\textbf {\bibinfo {volume} {5}},\
  \bibinfo {pages} {70} (\bibinfo {year} {2018})},\ \Eprint
  {http://arxiv.org/abs/1712.07109} {arXiv:1712.07109 [astro-ph.CO]}
  \BibitemShut {NoStop}%
\bibitem [{\citenamefont {Abi}\ \emph {et~al.}(2021)\citenamefont {Abi} \emph
  {et~al.}}]{Muong-2:2021ojo}%
  \BibitemOpen
  \bibfield  {author} {\bibinfo {author} {\bibfnamefont {B.}~\bibnamefont
  {Abi}} \emph {et~al.} (\bibinfo {collaboration} {Muon g-2}),\ }\href
  {\doibase 10.1103/PhysRevLett.126.141801} {\bibfield  {journal} {\bibinfo
  {journal} {Phys. Rev. Lett.}\ }\textbf {\bibinfo {volume} {126}},\ \bibinfo
  {pages} {141801} (\bibinfo {year} {2021})},\ \Eprint
  {http://arxiv.org/abs/2104.03281} {arXiv:2104.03281 [hep-ex]} \BibitemShut
  {NoStop}%
\bibitem [{\citenamefont {Gninenko}(2018)}]{Gninenko:2640930}%
  \BibitemOpen
  \bibfield  {author} {\bibinfo {author} {\bibfnamefont {S.}~\bibnamefont
  {Gninenko}} (\bibinfo {collaboration} {NA64 Collaboration}),\ }\href
  {http://cds.cern.ch/record/2640930} {\emph {\bibinfo {title} {{Addendum to
  the Proposal P348: Search~for~dark~sector~particles~weakly~coupled to muon
  with NA64$\mu$}}}},\ \bibinfo {type} {Tech. Rep.}\ \bibinfo {number}
  {CERN-SPSC-2018-024. SPSC-P-348-ADD-3}\ (\bibinfo  {institution} {CERN},\
  \bibinfo {address} {Geneva},\ \bibinfo {year} {2018})\BibitemShut {NoStop}%
\bibitem [{\citenamefont {Gninenko}(2019)}]{Gninenko:2653581}%
  \BibitemOpen
  \bibfield  {author} {\bibinfo {author} {\bibfnamefont {S.}~\bibnamefont
  {Gninenko}} (\bibinfo {collaboration} {NA64 Collaboration}),\ }\href
  {http://cds.cern.ch/record/2653581} {\emph {\bibinfo {title} {{ Proposal for
  an experiment to search for dark sector particles weakly coupled to muon at
  the SPS}}}},\ \bibinfo {type} {Tech. Rep.}\ \bibinfo {number}
  {CERN-SPSC-2019-002. SPSC-P-359}\ (\bibinfo  {institution} {CERN},\ \bibinfo
  {address} {Geneva},\ \bibinfo {year} {2019})\BibitemShut {NoStop}%
\bibitem [{\citenamefont {Kahn}\ \emph {et~al.}(2018)\citenamefont {Kahn},
  \citenamefont {Krnjaic}, \citenamefont {Tran},\ and\ \citenamefont
  {Whitbeck}}]{Kahn:2018cqs}%
  \BibitemOpen
  \bibfield  {author} {\bibinfo {author} {\bibfnamefont {Y.}~\bibnamefont
  {Kahn}}, \bibinfo {author} {\bibfnamefont {G.}~\bibnamefont {Krnjaic}},
  \bibinfo {author} {\bibfnamefont {N.}~\bibnamefont {Tran}}, \ and\ \bibinfo
  {author} {\bibfnamefont {A.}~\bibnamefont {Whitbeck}},\ }\href {\doibase
  10.1007/JHEP09(2018)153} {\bibfield  {journal} {\bibinfo  {journal} {JHEP}\
  }\textbf {\bibinfo {volume} {09}},\ \bibinfo {pages} {153} (\bibinfo {year}
  {2018})},\ \Eprint {http://arxiv.org/abs/1804.03144} {arXiv:1804.03144
  [hep-ph]} \BibitemShut {NoStop}%
\bibitem [{\citenamefont {Bode}\ \emph {et~al.}(2001)\citenamefont {Bode},
  \citenamefont {Ostriker},\ and\ \citenamefont {Turok}}]{Bode:2000gq}%
  \BibitemOpen
  \bibfield  {author} {\bibinfo {author} {\bibfnamefont {P.}~\bibnamefont
  {Bode}}, \bibinfo {author} {\bibfnamefont {J.~P.}\ \bibnamefont {Ostriker}},
  \ and\ \bibinfo {author} {\bibfnamefont {N.}~\bibnamefont {Turok}},\ }\href
  {\doibase 10.1086/321541} {\bibfield  {journal} {\bibinfo  {journal}
  {Astrophys. J.}\ }\textbf {\bibinfo {volume} {556}},\ \bibinfo {pages} {93}
  (\bibinfo {year} {2001})},\ \Eprint {http://arxiv.org/abs/astro-ph/0010389}
  {arXiv:astro-ph/0010389} \BibitemShut {NoStop}%
\bibitem [{\citenamefont {Fairbairn}(2022)}]{Fairbairn:2022gar}%
  \BibitemOpen
  \bibfield  {author} {\bibinfo {author} {\bibfnamefont {M.}~\bibnamefont
  {Fairbairn}},\ }\href {\doibase 10.3390/sym14040812} {\bibfield  {journal}
  {\bibinfo  {journal} {Symmetry}\ }\textbf {\bibinfo {volume} {14}},\ \bibinfo
  {pages} {812} (\bibinfo {year} {2022})}\BibitemShut {NoStop}%
\bibitem [{\citenamefont {Viel}\ \emph {et~al.}(2013)\citenamefont {Viel},
  \citenamefont {Becker}, \citenamefont {Bolton},\ and\ \citenamefont
  {Haehnelt}}]{PhysRevD.88.043502}%
  \BibitemOpen
  \bibfield  {author} {\bibinfo {author} {\bibfnamefont {M.}~\bibnamefont
  {Viel}}, \bibinfo {author} {\bibfnamefont {G.~D.}\ \bibnamefont {Becker}},
  \bibinfo {author} {\bibfnamefont {J.~S.}\ \bibnamefont {Bolton}}, \ and\
  \bibinfo {author} {\bibfnamefont {M.~G.}\ \bibnamefont {Haehnelt}},\ }\href
  {\doibase 10.1103/PhysRevD.88.043502} {\bibfield  {journal} {\bibinfo
  {journal} {Phys. Rev. D}\ }\textbf {\bibinfo {volume} {88}},\ \bibinfo
  {pages} {043502} (\bibinfo {year} {2013})}\BibitemShut {NoStop}%
\bibitem [{\citenamefont {Palanque-Delabrouille}\ \emph
  {et~al.}(2020)\citenamefont {Palanque-Delabrouille}, \citenamefont {Y\`eche},
  \citenamefont {Sch\"oneberg}, \citenamefont {Lesgourgues}, \citenamefont
  {Walther}, \citenamefont {Chabanier},\ and\ \citenamefont
  {Armengaud}}]{Palanque-Delabrouille:2019iyz}%
  \BibitemOpen
  \bibfield  {author} {\bibinfo {author} {\bibfnamefont {N.}~\bibnamefont
  {Palanque-Delabrouille}}, \bibinfo {author} {\bibfnamefont {C.}~\bibnamefont
  {Y\`eche}}, \bibinfo {author} {\bibfnamefont {N.}~\bibnamefont
  {Sch\"oneberg}}, \bibinfo {author} {\bibfnamefont {J.}~\bibnamefont
  {Lesgourgues}}, \bibinfo {author} {\bibfnamefont {M.}~\bibnamefont
  {Walther}}, \bibinfo {author} {\bibfnamefont {S.}~\bibnamefont {Chabanier}},
  \ and\ \bibinfo {author} {\bibfnamefont {E.}~\bibnamefont {Armengaud}},\
  }\href {\doibase 10.1088/1475-7516/2020/04/038} {\bibfield  {journal}
  {\bibinfo  {journal} {JCAP}\ }\textbf {\bibinfo {volume} {04}},\ \bibinfo
  {pages} {038} (\bibinfo {year} {2020})},\ \Eprint
  {http://arxiv.org/abs/1911.09073} {arXiv:1911.09073 [astro-ph.CO]}
  \BibitemShut {NoStop}%
\bibitem [{\citenamefont {Varma}\ \emph {et~al.}(2020)\citenamefont {Varma},
  \citenamefont {Fairbairn},\ and\ \citenamefont {Figueroa}}]{Varma:2020kbq}%
  \BibitemOpen
  \bibfield  {author} {\bibinfo {author} {\bibfnamefont {S.}~\bibnamefont
  {Varma}}, \bibinfo {author} {\bibfnamefont {M.}~\bibnamefont {Fairbairn}}, \
  and\ \bibinfo {author} {\bibfnamefont {J.}~\bibnamefont {Figueroa}},\
  }\href@noop {} {\  (\bibinfo {year} {2020})},\ \Eprint
  {http://arxiv.org/abs/2005.05353} {arXiv:2005.05353 [astro-ph.CO]}
  \BibitemShut {NoStop}%
\bibitem [{\citenamefont {He}\ \emph {et~al.}(2022)\citenamefont {He},
  \citenamefont {Robertson}, \citenamefont {Nightingale}, \citenamefont {Cole},
  \citenamefont {Frenk}, \citenamefont {Massey}, \citenamefont {Amvrosiadis},
  \citenamefont {Li}, \citenamefont {Cao},\ and\ \citenamefont
  {Etherington}}]{He:2020rkj}%
  \BibitemOpen
  \bibfield  {author} {\bibinfo {author} {\bibfnamefont {Q.}~\bibnamefont
  {He}}, \bibinfo {author} {\bibfnamefont {A.}~\bibnamefont {Robertson}},
  \bibinfo {author} {\bibfnamefont {J.}~\bibnamefont {Nightingale}}, \bibinfo
  {author} {\bibfnamefont {S.}~\bibnamefont {Cole}}, \bibinfo {author}
  {\bibfnamefont {C.~S.}\ \bibnamefont {Frenk}}, \bibinfo {author}
  {\bibfnamefont {R.}~\bibnamefont {Massey}}, \bibinfo {author} {\bibfnamefont
  {A.}~\bibnamefont {Amvrosiadis}}, \bibinfo {author} {\bibfnamefont
  {R.}~\bibnamefont {Li}}, \bibinfo {author} {\bibfnamefont {X.}~\bibnamefont
  {Cao}}, \ and\ \bibinfo {author} {\bibfnamefont {A.}~\bibnamefont
  {Etherington}},\ }\href {\doibase 10.1093/mnras/stac191} {\bibfield
  {journal} {\bibinfo  {journal} {Mon. Not. Roy. Astron. Soc.}\ }\textbf
  {\bibinfo {volume} {511}},\ \bibinfo {pages} {3046} (\bibinfo {year}
  {2022})},\ \Eprint {http://arxiv.org/abs/2010.13221} {arXiv:2010.13221
  [astro-ph.CO]} \BibitemShut {NoStop}%
\bibitem [{\citenamefont {Viel}\ \emph {et~al.}(2005)\citenamefont {Viel},
  \citenamefont {Lesgourgues}, \citenamefont {Haehnelt}, \citenamefont
  {Matarrese},\ and\ \citenamefont {Riotto}}]{Viel:2005qj}%
  \BibitemOpen
  \bibfield  {author} {\bibinfo {author} {\bibfnamefont {M.}~\bibnamefont
  {Viel}}, \bibinfo {author} {\bibfnamefont {J.}~\bibnamefont {Lesgourgues}},
  \bibinfo {author} {\bibfnamefont {M.~G.}\ \bibnamefont {Haehnelt}}, \bibinfo
  {author} {\bibfnamefont {S.}~\bibnamefont {Matarrese}}, \ and\ \bibinfo
  {author} {\bibfnamefont {A.}~\bibnamefont {Riotto}},\ }\href {\doibase
  10.1103/PhysRevD.71.063534} {\bibfield  {journal} {\bibinfo  {journal} {Phys.
  Rev. D}\ }\textbf {\bibinfo {volume} {71}},\ \bibinfo {pages} {063534}
  (\bibinfo {year} {2005})},\ \Eprint {http://arxiv.org/abs/astro-ph/0501562}
  {arXiv:astro-ph/0501562} \BibitemShut {NoStop}%
\bibitem [{\citenamefont {Garzilli}\ \emph {et~al.}(2021)\citenamefont
  {Garzilli}, \citenamefont {Magalich}, \citenamefont {Ruchayskiy},\ and\
  \citenamefont {Boyarsky}}]{Garzilli:2019qki}%
  \BibitemOpen
  \bibfield  {author} {\bibinfo {author} {\bibfnamefont {A.}~\bibnamefont
  {Garzilli}}, \bibinfo {author} {\bibfnamefont {A.}~\bibnamefont {Magalich}},
  \bibinfo {author} {\bibfnamefont {O.}~\bibnamefont {Ruchayskiy}}, \ and\
  \bibinfo {author} {\bibfnamefont {A.}~\bibnamefont {Boyarsky}},\ }\href
  {\doibase 10.1093/mnras/stab192} {\bibfield  {journal} {\bibinfo  {journal}
  {Mon. Not. Roy. Astron. Soc.}\ }\textbf {\bibinfo {volume} {502}},\ \bibinfo
  {pages} {2356} (\bibinfo {year} {2021})},\ \Eprint
  {http://arxiv.org/abs/1912.09397} {arXiv:1912.09397 [astro-ph.CO]}
  \BibitemShut {NoStop}%
\bibitem [{\citenamefont {Chandrasekhar}(1976)}]{10.2307/79115}%
  \BibitemOpen
  \bibfield  {author} {\bibinfo {author} {\bibfnamefont {S.}~\bibnamefont
  {Chandrasekhar}},\ }\href {http://www.jstor.org/stable/79115} {\bibfield
  {journal} {\bibinfo  {journal} {Proc. Roy. Soc. Lond. A}\ }\textbf {\bibinfo
  {volume} {348}},\ \bibinfo {pages} {39} (\bibinfo {year} {1976})}\BibitemShut
  {NoStop}%
\bibitem [{\citenamefont {Arbey}\ and\ \citenamefont
  {Auffinger}(2019)}]{Arbey:2019mbc}%
  \BibitemOpen
  \bibfield  {author} {\bibinfo {author} {\bibfnamefont {A.}~\bibnamefont
  {Arbey}}\ and\ \bibinfo {author} {\bibfnamefont {J.}~\bibnamefont
  {Auffinger}},\ }\href {\doibase 10.1140/epjc/s10052-019-7161-1} {\bibfield
  {journal} {\bibinfo  {journal} {Eur. Phys. J. C}\ }\textbf {\bibinfo {volume}
  {79}},\ \bibinfo {pages} {693} (\bibinfo {year} {2019})},\ \Eprint
  {http://arxiv.org/abs/1905.04268} {arXiv:1905.04268 [gr-qc]} \BibitemShut
  {NoStop}%
\bibitem [{\citenamefont {Akrami}\ \emph {et~al.}(2020)\citenamefont {Akrami}
  \emph {et~al.}}]{Akrami:2018odb}%
  \BibitemOpen
  \bibfield  {author} {\bibinfo {author} {\bibfnamefont {Y.}~\bibnamefont
  {Akrami}} \emph {et~al.} (\bibinfo {collaboration} {Planck}),\ }\href
  {\doibase 10.1051/0004-6361/201833887} {\bibfield  {journal} {\bibinfo
  {journal} {Astron. Astrophys.}\ }\textbf {\bibinfo {volume} {641}},\ \bibinfo
  {pages} {A10} (\bibinfo {year} {2020})},\ \Eprint
  {http://arxiv.org/abs/1807.06211} {arXiv:1807.06211 [astro-ph.CO]}
  \BibitemShut {NoStop}%
\bibitem [{\citenamefont {Carr}\ \emph {et~al.}(2010)\citenamefont {Carr},
  \citenamefont {Kohri}, \citenamefont {Sendouda},\ and\ \citenamefont
  {Yokoyama}}]{Carr:2009jm}%
  \BibitemOpen
  \bibfield  {author} {\bibinfo {author} {\bibfnamefont {B.~J.}\ \bibnamefont
  {Carr}}, \bibinfo {author} {\bibfnamefont {K.}~\bibnamefont {Kohri}},
  \bibinfo {author} {\bibfnamefont {Y.}~\bibnamefont {Sendouda}}, \ and\
  \bibinfo {author} {\bibfnamefont {J.}~\bibnamefont {Yokoyama}},\ }\href
  {\doibase 10.1103/PhysRevD.81.104019} {\bibfield  {journal} {\bibinfo
  {journal} {Phys. Rev. D}\ }\textbf {\bibinfo {volume} {81}},\ \bibinfo
  {pages} {104019} (\bibinfo {year} {2010})},\ \Eprint
  {http://arxiv.org/abs/0912.5297} {arXiv:0912.5297 [astro-ph.CO]} \BibitemShut
  {NoStop}%
\bibitem [{\citenamefont {Hasegawa}\ \emph {et~al.}(2019)\citenamefont
  {Hasegawa}, \citenamefont {Hiroshima}, \citenamefont {Kohri}, \citenamefont
  {Hansen}, \citenamefont {Tram},\ and\ \citenamefont
  {Hannestad}}]{Hasegawa:2019jsa}%
  \BibitemOpen
  \bibfield  {author} {\bibinfo {author} {\bibfnamefont {T.}~\bibnamefont
  {Hasegawa}}, \bibinfo {author} {\bibfnamefont {N.}~\bibnamefont {Hiroshima}},
  \bibinfo {author} {\bibfnamefont {K.}~\bibnamefont {Kohri}}, \bibinfo
  {author} {\bibfnamefont {R.~S.~L.}\ \bibnamefont {Hansen}}, \bibinfo {author}
  {\bibfnamefont {T.}~\bibnamefont {Tram}}, \ and\ \bibinfo {author}
  {\bibfnamefont {S.}~\bibnamefont {Hannestad}},\ }\href {\doibase
  10.1088/1475-7516/2019/12/012} {\bibfield  {journal} {\bibinfo  {journal}
  {JCAP}\ }\textbf {\bibinfo {volume} {12}},\ \bibinfo {pages} {012} (\bibinfo
  {year} {2019})},\ \Eprint {http://arxiv.org/abs/1908.10189} {arXiv:1908.10189
  [hep-ph]} \BibitemShut {NoStop}%
\bibitem [{\citenamefont {Keith}\ \emph {et~al.}(2020)\citenamefont {Keith},
  \citenamefont {Hooper}, \citenamefont {Blinov},\ and\ \citenamefont
  {McDermott}}]{Keith:2020jww}%
  \BibitemOpen
  \bibfield  {author} {\bibinfo {author} {\bibfnamefont {C.}~\bibnamefont
  {Keith}}, \bibinfo {author} {\bibfnamefont {D.}~\bibnamefont {Hooper}},
  \bibinfo {author} {\bibfnamefont {N.}~\bibnamefont {Blinov}}, \ and\ \bibinfo
  {author} {\bibfnamefont {S.~D.}\ \bibnamefont {McDermott}},\ }\href@noop {}
  {\  (\bibinfo {year} {2020})},\ \Eprint {http://arxiv.org/abs/2006.03608}
  {arXiv:2006.03608 [astro-ph.CO]} \BibitemShut {NoStop}%
\bibitem [{\citenamefont {Baur}\ \emph {et~al.}(2017)\citenamefont {Baur},
  \citenamefont {Palanque-Delabrouille}, \citenamefont {Yeche}, \citenamefont
  {Boyarsky}, \citenamefont {Ruchayskiy}, \citenamefont {Armengaud},\ and\
  \citenamefont {Lesgourgues}}]{Baur:2017stq}%
  \BibitemOpen
  \bibfield  {author} {\bibinfo {author} {\bibfnamefont {J.}~\bibnamefont
  {Baur}}, \bibinfo {author} {\bibfnamefont {N.}~\bibnamefont
  {Palanque-Delabrouille}}, \bibinfo {author} {\bibfnamefont {C.}~\bibnamefont
  {Yeche}}, \bibinfo {author} {\bibfnamefont {A.}~\bibnamefont {Boyarsky}},
  \bibinfo {author} {\bibfnamefont {O.}~\bibnamefont {Ruchayskiy}}, \bibinfo
  {author} {\bibfnamefont {E.}~\bibnamefont {Armengaud}}, \ and\ \bibinfo
  {author} {\bibfnamefont {J.}~\bibnamefont {Lesgourgues}},\ }\href {\doibase
  10.1088/1475-7516/2017/12/013} {\bibfield  {journal} {\bibinfo  {journal}
  {JCAP}\ }\textbf {\bibinfo {volume} {12}},\ \bibinfo {pages} {013} (\bibinfo
  {year} {2017})},\ \Eprint {http://arxiv.org/abs/1706.03118} {arXiv:1706.03118
  [astro-ph.CO]} \BibitemShut {NoStop}%
\end{thebibliography}%

\end{document}